\documentclass[preprint]{aastex}
\usepackage{emulateapj5}
\usepackage{amsmath}
\usepackage{natbib}

\begin{document}

\received{}
\accepted{}

\slugcomment{Version 10.0, a reduced version.  Full version\altaffilmark{0}
accepted by {\it The Astronomical Journal}.}
\lefthead{Peng et al.}
\righthead{Detailed Structural Decomposition of Galaxy Images}
\shortauthors{PENG ET AL.}
\shorttitle{DETAILED STRUCTURAL DECOMPOSITION OF GALAXY IMAGES}

\title{Detailed Structural Decomposition of Galaxy Images\altaffilmark{1}}

\author {Chien Y. Peng\altaffilmark{2}, Luis C. Ho\altaffilmark{3}, Chris D.
Impey\altaffilmark{2}, and Hans-Walter Rix\altaffilmark{4}}

\altaffiltext{0} {The full version of this paper with high resolution images
can be found at http://zwicky.as.arizona.edu/~cyp/work/galfit.ps.gz}

\altaffiltext{1} {Based on observations with the NASA/ESA {\it Hubble Space
Telescope}, obtained at the Space Telescope Science Institute, which is
operated by AURA, Inc., under NASA contract NAS5-26555.}

\altaffiltext{2}{Steward Observatory, University of Arizona, 933 N. Cherry
Av., Tucson, AZ 85721;  cyp, cimpey@as.arizona.edu.}

\altaffiltext{3} {The Observatories of the Carnegie Institution of Washington, 
813 Santa Barbara St., Pasadena, CA 91101;  lho@ociw.edu.}

\altaffiltext{4}{Max-Planck-Institut f\"{u}r Astronomie, Keonigstuhl 17,
	Heidelberg, D-69117, Germany;  rix@mpia-hd.mpg.de.}

\begin {abstract}

We present a two-dimensional (2-D) fitting algorithm (GALFIT) designed to
extract structural components from galaxy images, with emphasis on closely
modeling light profiles of spatially well-resolved, nearby galaxies observed
with the {\it Hubble Space Telescope}.  Our algorithm improves on previous
techniques in two areas, by being able to simultaneously fit a galaxy with an
arbitrary number of components, and with optimization in computation speed,
suited for working on large galaxy images.  We use 2-D models such as the
``Nuker'' law, the S\'ersic (de~Vaucouleurs) profile, an exponential disk, and
Gaussian or Moffat functions.  The azimuthal shapes are generalized ellipses
that can fit disky and boxy components.  Some potential applications of our
program include:  standard modeling of global galaxy profiles; extracting
bars, stellar disks, double nuclei, and compact nuclear sources; and measuring
absolute dust extinction or surface brightness fluctuations after removing the
galaxy model.  When examined in detail, we find that even simple-looking
galaxies generally require at least three components to be modeled accurately,
rather than the one or two components more often employed.  Many galaxies with
complex isophotes, ellipticity changes, and position-angle twists can be
modeled accurately in 2-D.  We illustrate this by way of 7 case studies,
which include regular and barred spiral galaxies, highly disky lenticular
galaxies, and elliptical galaxies displaying various levels of complexities.
A useful extension of this algorithm is to accurately extract nuclear point
sources in galaxies.  We compare 2-D and 1-D extraction techniques on
simulated images of galaxies having nuclear slopes with different degrees of
cuspiness, and we then illustrate the application of the program to several
examples of nearby galaxies with weak nuclei.

\end {abstract}

\keywords {galaxies: bulges --- galaxies: fundamental parameters --- galaxies: 
nuclei --- galaxies: structure --- technique: image processing --- technique: 
photometry} 

\section {INTRODUCTION}

Galaxies span a wide range of morphology and luminosity, and a very useful way
to quantify them is to fit their light distribution with parametric functions.
The de~Vaucouleurs $R^{1/4}$ and exponential disk functions became standard
functions to use after de~Vaucouleurs (1948) showed many elliptical galaxies
to have $R^{1/4}$ light distributions, while Freeman (1970) found later-type
galaxies to be well described by a de~Vaucouleurs bulge plus an exponential
disk.  Since then, the empirical techniques of galaxy fitting and
decomposition have led to a number of notable advances in understanding galaxy
formation and evolution.  These include investigations into the Tully-Fisher
relationship (Tully \& Fischer 1977), the fundamental plane of spheroids
(Faber et al.  1987; Dressler et al.  1987; Djorgovski \& Davis 1987; Bender,
Burstein, \& Faber 1992), the morphological transformation of galaxies in
cluster environments (e.g., Dressler 1980; van~Dokkum, \& Franx 2001), the
bimodality of galaxy nuclear cusps (Lauer et al. 1995; Faber et al.  1997) and
its implications for the formation of massive black holes (Ravindranath, Ho,
\& Filippenko 2002), and the cosmic evolution of galaxy morphology (e.g.,
Lilly et al. 1998; Marleau \& Simard 1998).

There are two general types of galaxy fitting: one-dimensional (1-D) fitting
of surface brightness profiles (e.g., Kormendy 1977; Burstein 1979; Boroson
1981; Kent 1985; Baggett, Baggett, \& Anderson 1998), and 2-D fitting of
galaxy images (e.g., Shaw \& Gilmore 1989; Byun \& Freeman 1995; de~Jong 1996;
Simard 1998; Wadadekar, Robbason, \& Kembhavi 1999; Khosroshahi, Wadadekar, \&
Kembhavi 2000), with each its own tradeoffs and benefits.

In 1-D, an important consideration is how to first obtain a radial surface
brightness profile from a 2-D image, for which there is no universally agreed
upon procedure.  A common practice is to use isophote fitting, which is a
powerful technique when performed on well-resolved galaxies because it
averages over elliptical annuli to increase the signal-to-noise ratio (S/N) at
a given radius.  However, as many galaxies have isophote twists and changing
ellipticity as a function of radius, the galaxy profile is extracted along a
radial arc that is ill-defined.  An alternative approach is to use a direct
1-D slice across an image.  Burstein (1979) argues that only cuts along the
major axis should be used in bulge-to-disk (B/D) decompositions.  Meanwhile,
Ferrarese et al.  (1994) point out that galaxies with power-law central
profiles may have different profiles along the major and minor axis.

Fitting profiles in 1-D is frequently used because it suffices for certain
goals, and is simple to implement.  But many studies now resort to 2-D
techniques.  For B/D decompositions, a number of authors (e.g., Byun \&
Freeman 1995; Wadadekar et al.  1999) have used idealized simulations to show
that 2-D modeling can better recover the true parameter values.  In 1-D, while
the galaxy bulge and disk may appear to merge smoothly, which causes
non-uniqueness in the decompositions, in 2-D, isophote twists and ellipticity
changes provide additional constraints to break those degeneracies.

While exponential and de~Vaucouleurs functions can model a wide range of
global galaxy profiles, galaxies are generally more complex.  Well resolved,
nearby galaxies are often poorly fit by the standard models in detail,
especially in 2-D.  Yet, few studies have gone beyond a global bulge and disk
decomposition, except in clear-cut cases where there are nuclear bars (e.g.,
de~Jong 1996) or nuclear point sources (e.g., Wadadekar et al. 1999; Portal
1999; Quillen et al. 2001; and many recent studies of quasar host galaxies).

More recently, a growing number of theories and {\it Hubble Space Telescope
(HST)} observations show that many clues of galaxy formation lay hidden in the
fine details of galaxy structure.  New {\it HST} images reveal striking
correlations between nuclear cusps of galaxies with their mass, stellar
velocity dispersion, radius, and other large-scale properties (Jaffe et al.
1994; Lauer et al.  1995; Faber et al.  1997; Carollo et al.  1997; Rest et
al.  2001; Ravindranath et al. 2001; Gebhardt et al. 2000; Ferrarese
\& Merritt 2000).  Those observations have led to on-going debates about
whether the correlations can be explained by an adiabatic growth of black
holes in isothermal cores and binary black hole mergers (e.g.  van der Marel
1999, and Ravindranath et al.  2002).  Other fossil remnants of galaxy
formation manifest as photometrically distinct nuclei, nuclear disks of stars
and gas, dust lanes, and nuclear spiral patterns (e.g., Phillips et al.  1996;
Carollo, Stiavelli, \& Mack 1998; Tomita et al. 2000; Tran et al.  2001).  If
elliptical galaxies form from 3:1 and 1:1 mass mergers, then Naab \& Burkert
(2001) merger simulations predict that all ellipticals should contain a
significant embedded stellar disk component in order to explain the kinematic
line profiles.  Is this seen, and if so, how often?  High resolution and high
S/N images now permit this to be tested.  Active galactic nuclei (AGNs) are
intimately related to formation of bulges and black holes; their fuel-starved
remnants are common tenants in nearby galaxies (Ho, Filippenko, \& Sargent
1997a).  Yet, quantitative investigations of these central sources have often
been hampered by the difficulties encountered in separating the faint nucleus
from the surrounding bright bulge (see, e.g., Ho 1999; Ho \& Peng 2001; Sarzi
et al.  2001).  In order to understand the physical implications of all these
clues, one must be able to extract accurate, quantitative morphological
information, via detailed galaxy decomposition.

Motivated by these and other possible applications, we developed a technique
(GALFIT) to accurately model galaxy profiles.  Section 2 discusses the
mechanics of our program and its computational requirements.  Section 3
provides an overview of our fitting procedures.  Section 4 presents a number
of case studies to illustrate the versatility of the program.  We then discuss
2-D decomposition to extract nuclear sources in \S~5, using simulations to
evaluate the recovery of point sources in 1-D compared to 2-D.  Then we apply
our method to some nearby galaxies that contain weak nuclear sources .
Conclusions are given in \S~6.  We make the source code readily available to
the public upon request.

\section {TWO-DIMENSIONAL FITTING PROGRAM}

We created a 2-D image decomposition program called GALFIT, written in the C
language.  In order to model galaxy profiles with a maximum degree of
flexibility, GALFIT uses a number of functions and can combine an arbitrary
number of them simultaneously.  To use the program the user provides a simple
input file, as shown in Figure~1.  In the example we specify a fit using five
functions simultaneously: a Nuker function, a S\'ersic, an exponential disk, a
Gaussian, and a uniform sky pedestal (see below for details).  Additional
components can be added by extending the list without limit, except by the
computer memory and speed available to the user.

\subsection {Accounting for Telescope and Atmospheric Seeing}

All recorded images carry an imprint of the observing set-up, due either to
imperfections or diffraction in the telescope optics, and sometimes by the
atmosphere.  To measure the intrinsic profile of an object, the seeing must be
taken into account either by convolution or deconvolution techniques.  Even
with {\it HST} WFPC2, whose optics are noted for near perfectly sharp,
diffraction-limited, point-source images, $10\%$ of stellar light falls
outside $0\farcs5$ of its core.  The point-spread function (PSF) of NICMOS on
{\it HST}\ has even more pronounced structure; the diffraction wings are
extended, and $\gtrsim 10\%$ of the flux lies beyond $\sim$$1\arcsec$.  One
way to remove the seeing is to deconvolve the image.  Lauer et al.\ (1995)
show that galaxy images with high S/N at the center can be accurately
deconvolved using $40-80$ iterations of Lucy-Richardson algorithm, if an
accurate PSF is available.  An alternative approach uses convolution, where
one constructs a model image and convolves it with a PSF before comparing the
result to data.  While both techniques rely on knowing an accurate PSF,
deconvolution has benefits when the S/N is high --- the observed image needs
only to be deconvolved once and one does not have to assume a particular model
for the intrinsic image.  However, deconvolution will not work well for
low-S/N or undersampled images because it may amplify Poisson and pixellation
noise.  In contrast, the convolution scheme works on images with both high or
relatively low S/N.  The drawback is that fitting an image proceeds more
slowly because convolution is done repeatedly, which is computationally
intensive.  Another disadvantage is that one needs to have reasonable a priori
knowledge of the input model.

GALFIT uses convolution, but it can be turned off if not needed.  Convolution
is done by using the convolution theorem:  we multiply the Fourier transforms
of the PSF and the models, and then inverse transform them.  Details of the
convolution and fitting processes are elaborated below.

\subsection {Computing the $\chi^2_\nu$}

The four input images for GALFIT are the CCD image of the galaxy, a noise
array, a PSF, and an optional dust (or bad pixel) mask --- all in FITS image
file format.  Pixels in the dust mask are rejected from the fit.

During the fit, $\chi^2_\nu$ (the reduced $\chi^2$) is minimized, defined in
the standard way as

\begin{equation}
\chi^2_\nu = \frac{1} {N_{\rm dof}}\sum_{x=1}^{nx}\sum_{y=1}^{ny} \frac {\left(\mbox{flux}_{x,y} - \mbox{model}_{x,y}\right)^2} {{\sigma_{x,y}}^2},
\end{equation}

\noindent where

\begin{equation}
\mbox{model}_{x,y} = \sum_{\nu=1}^{nf} f_{\nu,x,y}(\alpha_1...\alpha_n).
\end{equation}

\noindent $N_{\rm dof}$ is the number of degrees of freedom in the fit; $nx$
and $ny$ are the $x$ and $y$ image dimensions; and $\mbox{flux}_{x,y}$ is the
image flux at pixel $(x,y)$.  The $\mbox{model}_{x,y}$ is the sum of the $nf$
functions $f_{\nu,x,y}(\alpha_1...\alpha_n)$ employed, where
$\{\alpha_1...\alpha_n\}$ are the 2-D model parameters.  The uncertainty as a
function of pixel position, $\sigma_{x,y}$, is the Poisson error at each
pixel, which can be provided as an input image.  If no noise image is given,
one is generated based on the gain and read-noise parameters contained in the
image header.

\subsection {Galaxy and Compact Source Profiles}

\subsubsection {Azimuthal Shape and Pixel Sampling}

To fit galaxy and compact source profiles we assume the profiles all are
axially symmetric, generalized ellipses.  The form of the generalized ellipse
is motivated by Athanassoula et al.  (1990) who originally use it to describe
galaxy bar shapes.  When the principle axes of the ellipse are aligned with
the coordinate axes, the radial pixel coordinate is given by

\begin {equation}
r = \left(\left|x\right|^{c+2}+\left| \frac{y}{q}\right|^{c+2}\right)^{\frac{1}{c+2}}.
\end{equation}

\noindent An offset of 2 in the exponents is used so that a pure ellipse has
$c=0$, a boxy shape $c>0$, and a disky shape $c<0$.  This serves exactly the
same purpose as the cos~$4\theta$ Fourier coefficient frequently associated
with isophote fitting (Jedrzejewski 1987).  The parameter $c$ differs from the
cos~$4\theta$ Fourier coefficient in that the latter applies to individual
isophotes locally while the former applies to an entire component.  The
parameter $q$ is the ratio of the minor to major axis of an ellipse.  Both $q$
and $c$\ for each model are constant as a function of radius, although they
are free parameters.  Figure 2{\it a}\ and 2{\it b}\ show two examples of
azimuthal shapes as a function of $c$ for $q=1.0$ and $q=0.5$.  The PA of a
component is defined with respect to the image pixel coordinate system such
that a major axis positioned vertically is $0^\circ$ (nominally North if
rotated to the standard orientation) and increases counter-clockwise
(nominally toward the East).

When creating a model image, GALFIT decides whether or not to oversample the
pixels based on their distance from the centroid of a model component.  For
pixels far away from the center, the gradient is shallow; it suffices to
sample only at the center of each pixel.  But pixels near the center of a
component must be over-sampled.  For many cases dividing the pixels
up by square grids then summing the flux:

\begin{equation}
\mbox{model}_{\overline{x},\overline{y}} = \iint\limits_{pixel} f_{model} (x-\overline{x}, y-\overline{y})\mbox{ d}x\mbox{ d}y
\end{equation}

\noindent 

\noindent is sufficient.  However, for the Nuker profile (see below), when
$\gamma$ is large the function changes very quickly around $r=0$, and the
simple method quickly becomes inaccurate.  We devise a technique to integrate
the pixels around the center ($r<3$) by using an elliptical polar grid that
adapts to the ellipticity of the galaxy.  Figure~3{\it a}\ and 3{\it b}\
demonstrate the gridding for axis ratios of 0.9 and 0.3.  The angular spacing
is a 1-degree interval in a circular polar coordinate, modulo a tilt by the
inclination of the component.  The radial spacing of the grids increases
geometrically toward the center following

\begin {equation}
r_n = \frac{r_{n-1}}{(1 + \mbox{astep})^n}, 
\end {equation}

\noindent where $n$ increases closer to the center.  Empirically, astep = 0.01
works well for even the steepest Nuker profile, and we keep it as a
constant in the fit.

\subsubsection {Radial Profiles}

\noindent {\bf The S\'ersic Profile.} \ \ \ The S\'ersic (1968) profile has
the following form:

\begin {equation}
\Sigma(r)=\Sigma_e \mbox{ e}^{-\kappa \left[({\frac{r}{r_e}})^{1/n} - 1\right]}
\end {equation}

\noindent where $r_e$ is the effective radius of the galaxy, $\Sigma_e$ is the
surface brightness at $r_e$, $n$ is the power-law index, and $\kappa$ is
coupled to $n$ such that half of the total flux is always within $r_e$.  For
$n\gtrsim 2$, $\kappa\approx 2n - 0.331$; at low $n$, $\kappa(n)$ flattens out
towards 0 and is obtained by interpolation.  The original de~Vaucouleurs
(1948) profile is a special case with $n=4$ and $\kappa=7.67$.  While the
de~Vaucouleurs profile is well suited for ``classical'' bulges, some bulges
may be better represented by exponential profiles (e.g., Kormendy \& Bruzual
1978; Shaw \& Gilmore 1989; Kent, Dame, \& Fazio 1991; Andrekakis \& Sanders
1994).  The elegance of the S\'ersic profile is that it forms a continuous
sequence from a Gaussian ($n=0.5$) to an exponential ($n=1$) to a
de~Vaucouleurs ($n=4$) profile simply by varying the exponent.  It is very
useful for modeling bars and flat disks; the smaller the index $n$ is, the
faster the core flattens within $r<r_e$, and the steeper the intensity drop
beyond $r>r_e$.  The flux, integrated over all radii for an elliptical
S\'ersic profile with an axis ratio $q$ is

\begin{equation} 
F_{\rm tot} = 2\pi r_e^2 \Sigma_e e^{\kappa} n \kappa^{-2n}\Gamma(2n) q/R(c),
\end{equation}

\noindent where $\Gamma(2n)$ is the Gamma function. $F_{\rm tot}$ is converted
into a magnitude by GALFIT using the standard FITS exposure time parameter
(EXPTIME) in the image header.  $R(c)$ is a function that accounts for the
area ratio between a perfect ellipse and a generalized ellipse of
diskiness/boxiness parameter $c$, given by

\begin{equation}
R(c) = {\frac{\pi c} {4 \beta (1/c, 1+1/c)}},
\end{equation}

\noindent where $\beta(1/c, 1+1/c)$ is the ``Beta'' function with two
arguments.  In the 2-D implementation, the S\'ersic model has eight free
parameters: $x_{cent}, y_{cent}, F_{\rm tot}, r_e, n, c, q$, PA.  We note that
in place of fitting $\Sigma_e$, $F_{\rm tot}$ is fitted instead, which is more
often a useful parameter.  This is also done for all other models below,
except for the Nuker function.

\bigskip

\noindent {\bf The Exponential Disk Profile.} \ \ \ The exponential profile 
and the total flux are simply

\begin {equation}
\Sigma(r)=\Sigma_0 \mbox{ e}^{-\frac{r}{r_s}}
\end {equation}

\noindent
and

\begin {equation}
F_{\rm tot} = 2\pi r_s^2 \Sigma_0 q/R(c),
\end {equation}

\noindent where $\Sigma_0$ is the central surface brightness and $r_s$ is the
disk scale length.  The relationship between the half-light radius, $r_e$, and
the scale length, $r_s$ is $r_s = 1.678 r_e$ for this profile.  Most disky
galaxies are not composed of a single exponential disk, but also have either a
central S\'ersic or de~Vaucouleurs component.  They may also have either a
flat core or a truncated disk, which deviates from a simple exponential (e.g.,
van~der~Kruit 1979; Pohlen, Dettmar, \& L\"utticke 2000).

\bigskip

\noindent {\bf The Nuker Law.} \ \ \ The ``modified Nuker'' law was initially
proposed by Lauer et al.\ (1995) to fit the diverse inner 1-D profiles of 
galaxies observed with {\it HST}.  Its high degree of flexibility makes it
an excellent model for fitting most 1-D galaxy profiles.  The functional form 
of the Nuker law, which can be thought of as a double power law mediated by 
a smooth transition, is the following:

\begin{equation}
I(r) = I_b \ 2^{\frac{\beta - \gamma} {\alpha}}
\left({\frac{r}{r_b}}\right)^{-\gamma}\left[{1+\left(\frac{r}
{r_b}\right)^{\alpha}} \right] ^{\frac{\gamma-\beta}{\alpha}}
\end{equation}

\noindent This function has five adjustable parameters: $I_b, r_b, 
\alpha, \beta,$ and $\gamma$.  Taken to the limits of large and small radii,
the parameter $\gamma$ is the slope of the inner power law, and $\beta$ is the
slope of the outer power law.  The break radius $r_b$ is the location where
the profile changes slope, $I_b$ is the surface brightness at $r_b$, and
$\alpha$ describes how sharply the two power laws connect.  The more positive
$\alpha$ is, the sharper is the break at $r_b$.  In GALFIT, we fit the surface
brightness magnitude $\mu_b$ instead of $I_b$.  $\mu_b$ is obtained from $I_b$
via the use of the EXPTIME image header parameter and the pixel size specified
in item K of Fig.~1.  Although the profile appears singular at the center when
sampled at $r=0$, the integrated flux is finite for $\gamma \le 2$.  The 2-D
Nuker profile has a total of 10 free parameters:  $x_{cent}, y_{cent}, \mu_b,
r_b, \alpha, \beta, \gamma, c, q$, PA.

Although the parameters in Equation 1 correlate, the larger $\alpha$ is (i.e.
the sharper the break), the less coupled are $\alpha$, $\beta$, and $\gamma$,
when $r_b$ is well resolved (but much smaller than the image radius).  The
parameter coupling between $\beta$ and $\gamma$ is small for $\alpha$ fixed at
a value $\ge 1$, which is a useful reference point.  In this case, varying
$\beta$ across its whole range of typical values ($0<\beta<2.5$) would affect
the slope of $\gamma$ measured within $r\le0.1r_b$ by $\le 0.04$, for any
given value of $\gamma$.

\bigskip

\noindent {\bf The Gaussian Profile.} \ \ \ The Gaussian function and its
total flux are

\begin{equation}
\Sigma(r) = \Sigma_0 \mbox{ e}^{-\frac{r^2} {2 \sigma^2}}
\end{equation}

\noindent
and

\begin {equation}
F_{\rm tot} = 2\pi\sigma^2\Sigma_0 q/R(c).
\end {equation}

\noindent The full width at half maximum (FWHM) is 2.355$\sigma$.  The 2-D
model has seven free parameters: $x_{cent}, y_{cent}, F_{\rm tot}, \sigma, c,
q$, PA.

\bigskip

\noindent {\bf The Moffat/Lorentzian Profile.} \ \ \ The generalized Moffat
function has the following form,

\begin{equation}
\Sigma(r) = {\frac{\Sigma_0} {\left[1+(r/r_d)^2\right]^n}},
\end{equation}

\noindent
with the total flux given by 

\begin {equation}
F_{\rm tot} = {\frac {\Sigma_0 \pi r_d^2 q} {(n-1) R(c)}},
\end {equation}

\noindent where $r_d$ is the dispersion radius and $n$ is the power-law index.
The Moffat profile with $n$ of 1.5 or 2.5 is empirically similar in shape to
the observed WFPC2 PSF, while $n=1$ corresponds to a Lorentzian function.  It
has FWHM $= 2 r_d \sqrt{2^{1/n} - 1}$.  The Moffat model has eight free
parameters in 2-D.

\subsubsection {Creating Point Sources}

To create a true point source, one normally creates a $\delta$-function that is
then convolved with the input PSF.  But in GALFIT we approximate a
$\delta$-function by using a Gaussian (or Moffat) function having a small width,
usually FWHM $\lesssim 0.3$ pixel.  The benefit of using a functional
representation is that the algorithm can smoothly transit between fitting a
true point source or a compact source without a loss of generality.

\subsubsection {Creating Bars}

Spiral galaxies can have embedded bars, and to model them we prefer to use the
S\'ersic profile with initially a relatively flat inner and a steep outer
profile ($n<1$), and a boxy shape ($c>0$).  The true light distribution of a
bar also has a bulbous component at the center, distinct from the bulge.
Therefore one should in principle use two components.  However, the round
component may be partially degenerate with the bulge itself unless the two
profiles are significantly different.

\subsection{Computational Considerations}

One design consideration of GALFIT is to efficiently fit well-resolved
galaxies in large-format images, using an arbitrary number of model
components.  In our examples, the nearby galaxies extend beyond the Planetary
Camera (PC) chip of 800$\times$800 pixels.  As will be shown, they can be fit
accurately with typically three or more components, involving 20 or more free
parameters.  Fitting uncropped images is computationally intensive, and
convolving entire images can often usurp $> 99\%$ of the computing time and
memory.  However, in nearly all cases the seeing has significant impact only
near the central few arcseconds of a galaxy.  To make fitting large-format
images manageable, we convolve only the area most affected by the seeing
(Fig.~1, parameters H and I); the process is described below.  But the region
can be enlarged by the user as required.  We find that a convolution radius of
at least 20 to 30 seeing disks away from the center suffices for a wide range
of nuclear profile shapes.  For the {\it HST}\ WFPC2, this can correspond to a
region as small as $40\times40$ pixels.

In our implementation, computation time depends on the convolution size, the
number of parameters to fit, and the size of the fitted image.  Table 1 shows
the approximate resources used in running GALFIT for S\'ersic and exponential
disk fits where all parameters are free, as well as more complicated cases.
The total computation time required to reach convergence is estimated for an
Intel Pentium III 450 MHz computer with 128 megabytes of memory.  

\subsection {GALFIT Implementation}

We give an overview of the inner working of GALFIT to clarify the use of the
parameters shown in the top section of Figure~1.  Although we will go into
significant detail, we emphasize that the the actual fitting process is
completely transparent to the user, who needs only to prepare an input
template shown in Fig.~1.

GALFIT iterates steps 3-8 until convergence, with further explanations to
follow.

\begin {enumerate}
  \item Normalize and prepare the PSF for convolution (item D in Fig.~1).
  \item ``Cut out'' a section of the image centered on the object to fit from
        the original data image (item G in Fig.~1).
  \item Create model images and derivative images based on new or initial
	input parameters.
  \item ``Cut out'' the convolution region (items H \& I) from the model and
        derivative images in the previous step, and pad them around the edges
        with values of the models.
  \item Convolve the convolution regions (both model and derivative images) in
	previous step with the PSF using a Fast Fourier Transform (FFT)
	technique.
  \item Copy the convolution region back into the model/derivative images of
        step 3.
  \item Compare with data image. Minimization is done using the
	Levenberg-Marquardt downhill-gradient method/parabolic expansion
	(Press et al.\ 1997).
  \item Iterate from 3 until convergence is achieved.
  \item Output images and generate final parameter files.
\end {enumerate}

\bigskip

{\noindent {\bf The Input PSF.}} \ \ \ For convolution, GALFIT normalizes the
input PSF image (item D in Fig.~1) and rearranges it into a wrap-around order
(by splitting a PSF into four quadrants, then transposing them to diagonal
corners).  The PSF is then Fourier transformed using FFT.  In GALFIT, all
convolutions are done with FFT, a technique found in many numerical
computation texts (e.g., Press et al.\ 1997).  FFT reduces the computation
time by a factor proportional to $N^2/(N\hbox{ log}_2 N)$ compared to
brute-force numerical integrations, where $N$ is the number of pixels.

For {\it HST}\ observations, we obtain the PSF using either the Tiny Tim
(Krist \& Hook 1999) software or through the {\it HST}\ data archive.  For
WFPC2, the plate scale and PSFs are stable, and a number of studies find
that Tiny Tim can model the core structures of the PSFs, although the fine
details of the diffraction structures may be harder to reproduce.  The
encircled energy diagrams of the real and observed PSFs are very similar.  On
the other hand, data from the first year of NICMOS suffered from plate-scale
breathing and significant focus changes.  The small field of view (FOV) of
NICMOS ($20\arcsec-40\arcsec$) makes it difficult to find a PSF calibrator in
the same image for nearby galaxies.  However, Tiny Tim also allows the
modeling of time-varying plate-scale and breathing changes and can reproduce
the core structure of the PSF.  As with WFPC2, the diffraction rings and
structures are more difficult to reproduce.  For convolution and
deconvolution, the accuracy mostly depends on the encircled energy curves and
general structures of the diffraction pattern, and to a lesser extent on the
fine details.  Hence, the use of Tiny Tim PSFs should be adequate for
(de)convolution.  The user should supply a PSF large enough that the flux
amplitude around the image edges is negligible compared to the peak ($\ll
1\%$); any sky pedestal must also be removed in the PSF image.

\bigskip

{\noindent {\bf Extracting a Sub-Image.}} \ \ \ GALFIT can either fit the
entire image or a sub-region (item G in Fig.~1), allowing reduced computation
time for a small object.  This is also useful when one might want to fit the
nucleus of a galaxy accurately, before enlarging the fit to outer regions.

\bigskip

{\noindent {\bf Creating Model and Derivative Images.}} \ \ \ To optimize the
galaxy profile parameters, GALFIT uses a down-hill gradient method.  Unlike
most other methods which create only model images, the down-hill gradient
technique requires both the model and flux derivative images with respect to
all free parameters at each pixel.  Because they are stored in memory, the
memory usage scales as $N_{\rm free} \times N_{\rm pix} + 1$, where $N_{\rm
free}$ is the number of free parameters in the fit and $N_{\rm pix}$ is the
number of pixels in an image.

\bigskip

{\noindent {\bf Extracting the Convolution Region.}} \ \ \ Once the model and
derivative images are created, they are convolved with the PSF to account for
the seeing.  In order to minimize the computation time and memory in the
convolution, GALFIT removes a small region (specified in items H and I in
Fig.~1) from the model and extends it with additional padding.  Padding is
needed because of edge effects resulting from convolution (see Press et al.\
1997).  GALFIT performs two padding operations, first by half the size of the
PSF all around the user-specified convolution region.  That region is filled
with {\it real\ } model values rather than with zeros --- padding with zeros
would also corrupt the model around the edges into the convolution region.
The image is then further extended with zero padding up to the next $2^N$ ($N$
integer) number of pixels in both width and length.  This is the actual image
size used in the convolution.  This process is performed for all the models,
as well as the derivative images.

\bigskip

{\noindent {\bf Convolving with the PSF.}} \ \ \ The convolution regions
removed from the model and derivative images are convolved with the PSF using
FFT.  If a PSF is not specified, as when an image has already been
deconvolved, the convolution step is bypassed.

\bigskip

{\noindent {\bf Copying the Convolution Region back to the Model and
Derivative Images.}} \ \ \ Once convolution is complete, the model/derivative
images are copied back into the original model/derivative images.  The two
surrounding layers of (corrupted) padding regions are discarded.

\bigskip

{\noindent {\bf Minimizing Residuals.}} \ \ \ To minimize residuals, we choose
the Levenberg-Marquardt method (Press et al.\ 1997) as the engine instead of
the Simulated Annealing algorithm.  We discuss our decision in \S~3.1.  The
process of minimization repeats until convergence is achieved, which happens,
artificially, when the $\chi^2$ does not change more than 5 parts in $10^4$
for 5 iterations.

In complicated fits as the examples we will perform, there is a non-zero
chance that GALFIT will fail to produce a good solution.  One kind of failure
is when the program quits (``crashes'') without a solution because the
solution matrix is singular, either caused by poor parameter values, or trying
to fit more components into the image than appropriate.  GALFIT is generally
highly forgiving about poor initial values and frequently converges to good
fits ($\chi^2_{\nu} \approx 1$-2), down from $\chi^2_\nu$ as high as
$10^5$-$10^7$.  When GALFIT does crash, more often than not it is caused by
numerical overflows when the the power law indices ($\alpha$ and $\gamma$ for
Nuker, and $n$ for S\'ersic) become too big ($> 10$) or too small ($\lesssim
0.01$).  Another kind of failure is when the solution settles into a local
$\chi^2$ minimum and does not get out, but this will not cause GALFIT to
crash.  For example, in B/D decompositions, there are correlations between the
scale-length and luminosity parameters (see, e.g., Byun \& Freeman 1995),
while ellipticities, centers, and position angles are generally more decoupled
and better constrained by the data.  We discuss the issue of degeneracy in
more detail in \S~3.3.

\bigskip

{\noindent {\bf Output Images.}} \ \ \ Upon completion of the minimization, a
FITS image block is created, consisting of three images:  the original, the
model, and the residual.  The software provides the option to not subtract a
fitted profile, thereby leaving it in the residual image (item Z).

\subsection{Parameter Uncertainty Estimation}

Once a fit has been optimized, GALFIT estimates the uncertainties
analytically.  We assume that the surface of constant $\Delta\chi^2$, as a
function of {\it n} free parameters, away from the minimum $\chi^2$ can be
approximated by an {\it n}-dimensional ellipsoid.  This allows the
uncertainties to be obtained from the covariance matrix, defined as the
local curvature of the $\chi^2$ surface at $\chi^2_{min}$ with respect to each
parameter such that $\mbox{C}_{ij} =\mbox{d}^2\chi^2/(\mbox{d}a_i\
\mbox{d}a_j)$, where $a_i$ and $a_j$ are the fitted parameters.  The standard
68\% bound on the confidence interval of the fitted parameters, individually,
is within a boundary marked by a shell at $\Delta\chi^2=1$ (e.g. Press et al.
1997).  Figure 4 shows an example for a 2-parameter ($a_1$ and $a_2$) fit.

When the parameters are uncorrelated, the off-diagonal entries of the
covariance matrix are zero, and the ellipsoid has major axes running parallel
to the unit coordinate axes $\hat{a}$.  Formally, the parameter uncertainties
are related to the eigenvalues of the covariance matrix, and the standard
68\%, 1-$\sigma$ uncertainty is $\sigma a_i =
\sqrt{2/\mbox{C}_{ii}}$.

However, when some of the parameters are correlated, and the $\Delta\chi^2$
ellipsoid axes are no longer aligned with the parameter axes.  Figure~4 shows
an example for a 2-parameter ($a_1$ and $a_2$) fit, where $a_1$ and $a_2$ are
correlated.  Generalizing to higher dimensions, to estimate uncertainties in
one parameter we first obtain the eigenvectors and eigenvalues of the {\it
n}-dimensional hyper-ellipsoid.  The semi-major axis vector $i$ is given by:

\begin{equation}
\vec{v}_i = {\sqrt{\frac{\Delta\chi^2}{\lambda_i}}} \hat{v}_i, 
\end{equation}

\noindent where $\lambda_i$ is the eigenvalue corresponding to the eigenvector
$\hat{v}_i$.  We define the uncertainty for parameter $a_i$ by vector summing
all the semi-major axis vectors in such a way that the projection onto a given
axis $\hat{a}_i$ is largest.  In Figure 4, these are vectors represented by
the arrow-tipped dotted lines.  We then multiply this quantity by a factor of
$\sqrt{\chi^2_\nu}$, i.e. we renormalize the errors, to avoid underestimating
the errors when $\chi^2_\nu$ is not ideally 1.  Written explicitly,

\begin{equation}
\sigma a_i = \sqrt{\chi^2_\nu} \cdot \sum_{j=1}^{n}
|\vec{v}_j\cdot{\hat{a}}_i|.
\end{equation}

\noindent Strictly speaking, our definition of the uncertainty in Equation 17
is equal to, or slightly larger than, the true uncertainties (which are
$\sigma a_1\arcmin$ and $\sigma a_2\arcmin$ in Figure 4).  We settle for this
possible overestimate, but do not consider it to be serious, because the
fundamental limitation in determining precise errors is that we do not know
how the $\chi^2$ ``valley'' may twist near the best fit $\chi^2$ or how well
it is represented by a hyper-ellipsoid.  Furthermore, the parameters we fit
may not be all normally distributed.  We verified using artificial data with
Poisson noise that our method produces reasonable uncertainty estimates for
simple models consisting of one and two input components.

Uncertainty estimates quoted in the text and table below are obtained using
the method described unless otherwise stated.

\section {GALAXY FITTING: GENERAL CONSIDERATIONS}

\subsection{The Choice of the Minimization Algorithm}

The uncertainty estimate just described is only valid when the solution has
reached a global minimum.  However, the $\chi^2$ topology for galaxy fitting
is complex with many local minima (degeneracies) because the parameter space
is large (e.g., $\geq16$ parameters for two or more S\'ersic components).
While no algorithm can guarantee convergence on the global minimum $\chi^2$,
sophisticated Simulated Annealing (also known as Metropolis or Annealing for
short) algorithms are often the engine of choice for automating B/D
decompositions, where they are reputed to be robust.  In contrast, though
vastly more efficient, the gradient method may head downhill blindly,
regardless of whether it is going toward a global $\chi^2$ minimum.  Despite
this shortcoming, we use the downhill-gradient/parabolic expansion method of
Levenberg-Marquardt (Press et al. 1997) as our minimization engine, for
reasons given below.

An algorithm that uses Simulated Annealing is GIM2D, developed by Simard
(1998) for two-component B/D decomposition.  In this method, the annealing
(i.e., cooling) ``temperature'' step size controls the rate of convergence on
a solution.  At each iteration, a set of parameters is perturbed randomly by
some amount; high temperatures correspond to large perturbations.  Then, the
probability $P_1$ that the parameter set is the true one is calculated.  If
this new probability is higher than the previous, $P_0$, the new parameter set
is adopted.  However if the new set is less likely, there is nonetheless a
finite chance $P_1/P_0$ that it is adopted.  At some point the temperature is
decreased and the iteration continues until convergence.  The application of
the software is demonstrated in several studies (e.g., Marleau \& Simard 1998;
Simard 1998).

Fitting more than two components dramatically increases the complexity of the
$\chi^2$ topology, and extending the Annealing technique to fit more
components would seem to be sensical.  However, in practice it is tricky to
find an optimum cooling rate that can generally accommodate the different
number, and type, of components a galaxy might require.  Anneal too quickly,
the solution settles into a local $\chi^2$ minimum; too slowly, the
program converges even more inefficiently.  Efficiency is an issue because
increasing the number of parameters makes it exponentially more cumbersome
to adequately sample the parameter space for calculating probabilities.  As an
extreme example, to fit the double nucleus of M31 accurately, one needs 6
components, having 41 to 50 free parameters (Peng 2002).  Moreover, for highly
resolved nearby galaxies, one generally has to create dust masks and test out
the components iteratively.  This occasional need for interaction makes
both the speed and the straightforward implementation of the downhill-gradient
method attractive.

The downhill-gradient method may not be quite as well optimized for B/D
decomposition as the Simulated Annealing method.  However, a direct comparison
between the two methods is needed to evaluate their respective strengths and
weaknesses.  We discuss below (\S~3.3) how to test for degeneracies.  It may
also be possible to tailor a hybrid gradient-search and Annealing method,
which tries to climb out from a local minimum by briefly using Monte-Carlo
sampling after the solution has been optimized by the gradient search.  This
is being considered for future updates to GALFIT.

\subsection {Fitting Procedures}  

Most bulges and late-type galaxy nuclei that are well resolved by {\it HST}\
appear as smooth and simple spheroids on large scales when one disregards the
occasional dust.  But, as we will show below, simple one or two component fits
frequently produce large residuals that have bipolar or quadrupolar symmetry,
which can be reduced with additional components.  This subtlety makes
estimating the required number of components a trial-and-error process, and
Occam's razor suggests using the fewest number necessary, based on the goal of
the experiment.  We determine the number and types of components iteratively,
often by starting out with a S\'ersic or a Nuker profile for ellipticals, and
a S\'ersic plus an exponential for spirals.  Based on $\chi^2$ and the pattern
of residuals, we determine if we need to replace them with more flexible
models or add in more components.

To use GALFIT in a typical manner such as for B/D decomposition, we point out
that there is little to no human interaction beyond preparing an initial
parameter file.  However, our task of accurately fitting nearby large galaxies
is considerably more challenging, thus we outline the steps we normally take:

\begin{enumerate} 

  \item Create a dust/bad pixel mask by hand.  We create a mask by outlining
	the affected region with a polygon, then feed the list of pixels
	into GALFIT either as a text file or a FITS image.
  \item Estimate by eye the number of components needed.
  \item Although not needed, it sometimes helps to initially hold the
	diskiness/boxiness parameter $c$ fixed until the algorithm has found a
	plausible solution.  The radicands in the radius Equation [1] have
	absolute value quantities; hence, the derivative images contain
	regions of discontinuity that cause the downhill-gradient method to
	react sensitively unless near a minimum.
  \item If convolution is needed, leaving it off initially saves
        time until a reasonable solution has been attained.
  \item If a galaxy is difficult to fit near the center, we restrict fitting
	to that region until a good solution is found, then enlarge
	the fit.
  \item Optimize all the parameters in a $\chi^2$ sense.
  \item Examine the residuals to decide whether more components need to be 
        added, make better pixel masks, free parameter values that have
        first been held fixed, or turn on convolution.  
  \item Repeat previous steps as necessary.
\end{enumerate}

\subsection{Degeneracy and the Significance of the Components}

In this paper and in other related studies (Peng 2002; Ho et al. 2002), we
find that galaxy light profiles can generally be modeled accurately (in a
$\chi^2$ sense) using three to five components.  The larger-than-usual number
of components immediately prompts the question of how unique is the
decomposition.

Degeneracy is a common problem in galaxy decompositions, mostly because the
model functions used are not selected based on physical criteria.  One can, in
principle, decompose a galaxy with as many ``basis functions'' as one might
contrive.  Even when a set is well defined, such as in two-component B/D
decompositions, the large number of parameters involved makes degeneracy
sometimes an issue, as shown in the simulations of Byun \& Freeman (1995) and
Wadadekar et al.  (1999).  Extending to more components might, at first sight,
make the situation unmanageable.  One way to reinforce our confidence in {\it
parameter}\ uniqueness is to use Monte-Carlo simulations to search the
parameter space, but these say nothing about whether a given {\it combination
of models}\ is the appropriate or unique set.  Moreover, searching randomly
may be unfeasible because the dimensionality is over 20, and occasionally over
40.  However, while there are potentially many local $\chi^2$ minima
solutions, in practice the situation is not hopelessly degenerate.  Imposing
the demand that models fit galaxies with high fidelity ($\chi^2_{\nu}\approx
1$) severely reduces the possible solution space for a given set of models;
most solutions can be easily dismissed based on $\chi^2$ analysis.  Such
demand produces components that have significantly different scale lengths,
axis ratios, etc..

While there is no recipe to guarantee uniqueness in the fitting, there are
ways to probe the parameter spaces exhaustively and to satisfy to one's own
confidence that a solution is stable and insensitive to initial conditions.
One way is to vary two parameters at a time on a Cartesian grid of values in
order to trace out the local $\chi^2$ contours.  This technique is very useful
when the number of parameters is low.  However, the same technique quickly
becomes intractable with increasing number of fitting parameters.  To more
effectively explore a wider $\chi^2$ topology, we propose the following 
technique of combining Monte-Carlo simulations and minimization.

\begin{enumerate}
  \item Randomly select a full set of parameters, possibly drawing from
	distributions of parameter values centered on the best-fit values.
  \item Using these as initial values, minimize the $\chi^2$.  
  \item Repeat step 1 for many different sets of random initial guesses to
        see if they return the same optimized solution or to other
	equally plausible ones.
\end {enumerate}

The degeneracy situation, together with the use of more than three components,
makes interpretation difficult, leading one to ask: ``what do all the
components mean?'' or ``why use so many components?''  The answer depends on
the prior goal:  Perhaps the most mundane answer is simply that a bulge is
triaxial, or otherwise not well represented by our limited models.  In this
case two or more functions may be needed to describe the same entity that has
meaning only when summed, but not individually.  A different, but perhaps
more interesting tactic is to explicitly seek out components motivated by
theory.  A case in point is the prediction by Naab \& Burkert (2001) that
there should be disks embedded in elliptical galaxies formed in galaxy
mergers. The use of extra components may also be motivated by external data,
such as evidence for a nuclear point source based on AGN spectra, or decoupled
cores based on kinematic data.  To look for weak AGN point sources, it is
especially important to deblend a large portion of an image accurately, rather
than biasing the fit to regions that one can fit well.  Lastly, from a purely
empirical standpoint, one can decompose a galaxy simply to look for structures
too subtle to be seen in full light, such as weak nuclear bars, stellar disks,
and nuclear point sources.  Again, accurate decomposition here is crucial so
that the residuals do not undermine the detection or believability of weak
sources.

The last scenario brings the discussion back in full circle to the issue of
degeneracy.  Despite possible degeneracies in decomposing large-scale
components, smaller structures are better localized, and better defined in
size and shape.  Even if there is doubt as to what the exact profile and
parameters are, often different models can reproduce a component with similar
overall characteristics (e.g., profile, scale-length, shape, and orientation),
even if they are not exactly identical.  That a similar component is
repeatedly borne out through different profile assumptions is in principle the
keystone of its reality, since its inclusion is essential to achieving a
reasonable model fit in that region of the image.  We show a few such cases
below.

\section {MODEL FITTING: A CASE BY CASE STUDY}

In the discussion to follow, we present GALFIT decomposition of galaxies
having a wide variety of properties and shapes in order to test how well
nearby galaxies can be modeled in 2-D by parameterized functions.  We apply
GALFIT to images of 7 galaxies that have a wide range of interesting
properties: different shapes, morphologies, isophote twists, and internal
disks.  Our WFPC2 data come from the {\it HST}\ archive.  We remove the edges
which are affected by CCD readouts, and fit models to $721\times721$ pixel
images.  The galaxy nuclei in these images have very high S/N and are
centrally placed in the PC.  They often extend well beyond the WFPC2 FOV, and
the exposures are short (several hundred seconds, see Table 1), making the sky
background and image bias level hard to estimate from the Wide Field chips.
Given the typical exposure times of our images, the 26 mag arcsec$^{-2}$
isophote is about 0.01 -- 0.05 ADU.  We set the sky to zero in our fitting,
unless specifically mentioned otherwise.

Figures $5-10$ and Figure $14$ show the original images as well as the
residuals in positive greyscale.  We show two panels of residuals.  The first
(panel {\it b}) illustrates the traditional decomposition technique using only
a one-component bulge and disk, occasionally adding a bar and a point source
when obviously needed.  We use a S\'ersic or Nuker function instead of de
Vaucouleurs model for the bulge.  The second residual panel (panel {\it c})
shows a more accurate decomposition using the components listed in Table 3.

Beneath the images are plots of the surface brightness profile, PA,
ellipticity, and cos~$4\theta$ obtained by running the IRAF\footnote{IRAF
(Image Reduction and Analysis Facility) is distributed by the National Optical
Astronomy Observatories, which are operated by AURA, Inc., under cooperative
agreement with the National Science Foundation.} task {\it ellipse}.  The
surface brightness plot shows the {\it observed}\ profile (i.e., without
deconvolution) as solid circles with error bars.  The short-dashed line going
through those points is a 1-D Nuker law that best fits the observed profile;
this is merely to illustrate the flexibility of the function, as well as to
guide the eye for deviations from smoothness or power-law behavior.  The
long-dashed lines show the {\it intrinsic}\ profile of each component listed
in Table 3 from 2-D decomposition.  Finally, the solid line is the net sum of
all the components that make up the intrinsic profile.  Therefore, it will
rise above the observed points at the nucleus in cases when there is no point
source.  The dotted vertical line in these plots shows the region most
affected by PSF smearing, but in some cases the effect can be seen as far out
as $0\farcs6$ (e.g., NGC 221).

The ellipticity is defined as $\epsilon=1-q$, where $q$ is the axis ratio,
and the cos~$4\theta$ plot shows whether the isophotes are disky ($>0$) or
boxy ($<0$), similar in definition to the $c$ parameter in GALFIT, but
opposite in sign.

Table 3 summarizes all the fitting parameters; the magnitudes have not been
corrected for Galactic extinction.  The magnitudes are based on the synthetic
zeropoint calibrations by Holtzman et al. (1995), which transform data units
onto the Landolt UBVRI magnitude system.  Table 3, column 15, lists the
$\chi^2_\nu$ for the best fit shown in Figures {\it c}, and column 16 lists
the increase in $\chi^2_\nu$ for fits shown in Figures {\it b}.  We do not
quote uncertainties for all the parameters individually.  Instead, in the last
row of Table 3, we quote representative uncertainties derived in the fit (see
\S~ 2.6) as an ensemble.  For most fits, ``representative'' means a
conservative upper bound on the fitting parameters.

\subsection {A Caveat on Galaxy Morphology Comparisons}

Comparing galaxy morphology/decompositions using inhomogeneous studies is
difficult if they are based on images of limited FOV and S/N.  Galaxies in
low-S/N images have wings that are barely detectable above the sky noise,
often biasing scale-length measurements toward lower values, if the noise is
not accurately accounted or the suitable function is unknown.  While low S/N
in the wings is offset by a larger number of pixels, there is a practical
tradeoff between areal increase and the rate of profile decline in the
presence of sky and detector noise.  In decomposing nearby galaxies there is
an additional concern if they extend well beyond the FOV of the CCD.  Since
galaxy profiles tend to change shape, PA, and ellipticity with radius, in
addition to the degeneracies discussed above, it is generally unsafe to
extrapolate profiles beyond the FOV.

The galaxies in our examples below were all observed with WFPC2.  And although
they have very high S/N in the core, they have short exposure times (several
hundred seconds).  Moreover, they show only the central regions of the
galaxies.  While they are perfectly suited for studying galaxy nuclei, profile
comparisons with larger FOV images are not straightforward.  When we fit
components whose scale-lengths are comparable to, or larger than, the FOV, we
do not necessarily believe the numbers to be accurate.  However, the reality
of such a component is often suggested by analysis of the residuals.

We leave further interpretations to future studies (Ho et al.\ 2002).

\subsection {A ``Simple'' Elliptical, NGC 221 (M32)}

This E2 galaxy has an apparent axis ratio of 0.72.  Michard and Nieto (1991)
find that M32 has pointed isophotes between 1\farcs2 and 5$\arcsec$.  Color
images obtained with {\it HST}\ and presented in Lauer et al.\ (1998) show
strong surface brightness fluctuations, which were studied by Ajhar et al.\
(1997).  The surface photometry of M32 is shown in Figure~5, uncorrected for
PSF smearing.  Lauer et al.\ (1992, 1998), through deconvolution of WFPC and
WFPC2 data, find that the ellipticity is constant throughout the inner
portions of the galaxy.  They see no signs of significantly boxy or disky
isophotes, nor evidence for dust, disks, or other structure down to a few
percent in local surface brightness.

In 1-D, the Nuker function fits the surface brightness profile remarkably well
out to $r\,\approx\,20\arcsec$.  However, a 2-D Nuker fit leaves behind large
residuals near the center (Fig.~5{\it b}).  We need two components (see Table
3) to produce a good fit (Fig.~5{\it c}).  The requirement of two components
is not due to ellipticity and PA changes; in fact, in agreement with Lauer et
al.  (1998), the PA, ellipticity, and diskiness/boxiness parameters all appear
to be constant with radius in our two components.  Rather, two components are
needed because of a subtle upturn in the profile in the region
$0\farcs7-2\arcsec$.  This is nearly indiscernible in the 1-D surface
brightness profile of Figure~5, but much more so in Figure~5{\it b}\ where a
one-component Nuker fit leaves a ``doughnut-shaped'' residual, with an
increase in $\chi^2_{\nu}$ by 0.13 from the best fit, Fig.~5{\it c}.

Much of the structure can be removed by fitting two components to the image.
Figure~5{\it c}\ shows a S\'ersic + exponential disk fit.  A reasonable fit
can also be obtained with a de~Vaucouleurs bulge plus a nuclear disk described
by Nuker profile.  We adopt the first set of parameters because formally it is
better.  Despite the excellent fit, the residuals show a very low diffuse
positive halo out to $1\farcs3$, at a level about $0.1\%-1\%$ of the local
flux.  The need for two components point to the intriguing possibility that
there might be a nuclear stellar disk with a semi-major axis of
$r_s\approx1.5$ pc.  There are also slight negative residuals around
($20\arcsec,18\arcsec$), which may suggest the presence of dust at a level
$\sim 3\%-8\%$ below the local mean intensity, which is not seen in the region
reflected about the major axis.  Hence, this is probably not caused by a
profile mismatch because mismatches would produce bipole or quadrupole
patterns.  We confirm Michard \& Nieto's (1991) finding that M32 has a slight
disky component, as evidenced in both Figure~5 and in our $c$ parameter
($c=-0.06\pm0.01$).

\subsection {``Disky'' Galaxies}

\subsubsection {NGC 4111}

This nearby edge-on S0 galaxy is distinguished by highly disky isophotes, a 
peanut-shaped nucleus (pinched at the minor axis), and a complex
surface brightness profile (Fig.~6) having multiple breaks and inflections
at large radii.  Burstein (1979) and Tsikoudi (1980) analyze photographic
plate scans of NGC 4111 that cover 1$^{\prime}$--4$^{\prime}$ in radius.  Upon
decomposing it in 1-D, they find several distinct photometric components.
Tsikoudi (1980) finds that the inner spheroidal component with a classic
$R^{1/4}$ profile has an effective radius $r_e = 15\arcsec$ along the
semi-major axis; it contributes 48$\%$ of the total luminosity in the $B$
band.  The remainder of the galaxy luminosity is contained in the ``lens''
component of the galaxy and in an exponential disk that extends beyond
$102\arcsec$.  There are several profile inflections at large radii, in
particular at $a\approx30\arcsec$ and $70\arcsec$; Tsikoudi suggests that the
hump between $30\arcsec < a < 102\arcsec$ may be a ring or weak spiral arm.
Cross-cuts perpendicular to the major axis (perpendicular profile) at
$10\arcsec$, $30\arcsec$, and $50\arcsec$ from the center show that the
exponential envelope itself is composed of three components, where the inner
(which extends out to 12$\arcsec$) and intermediate regions ($<24\arcsec$) are
Gaussians, while the outer region is characterized by an exponential disk with
a steep drop-off, extending well beyond $60\arcsec$ (Tsikoudi 1980).
Similarly, Burstein finds that the perpendicular profiles of NGC 4111
can be described by an exponential disk profile, and that the rounded center 
can be fitted by a bulge component.  Moreover, the residuals suggest there is a
``thick disk'' component, which is also seen in other S0s in his sample.

The innermost 20$\arcsec$ of NGC 4111, as seen with {\it HST}, proves to be
just as complicated as the region exterior, with several breaks in the
light profile (Fig.~6).  2-D decomposition reveals that much of the dip at
$r=1\arcsec$ is caused by a circumnuclear dust ring.  The dust lane bisecting
the nucleus is not fully symmetric in shape, and gives rise to the peanut
appearance.  The two-component B/D decomposition (Nuker and exponential)
presented in Figure~6{\it b} shows large over- and under-subtractions.  There
is a prominent disky component with a boxy nucleus in the residuals.
Moreover, at the junction between the two, the disk appears to be pinched.

A more accurate decomposition of the nuclear region requires four disky
components and a boxy component.  The five components give an excellent fit to
the nucleus, despite all the complexities, and we claim that at least four are
real and photometrically distinct.  Of the two dominant components, one is an
exponential disk ($r_s=21\farcs4$) with a disky shape ($c=-0.65$).  The other
is a S\'ersic ($r_e=7\farcs85$) with $n$ = 2.51 and a boxy shape ($c=0.28$),
giving rise to the peanut-bulge appearance.  There is also a nuclear stellar
disk that stands out very prominently in the decomposition.  It consists of
two components that have nearly identical half-light radius of $r_e=
4\farcs5$: a Gaussian halo ($n=0.48$) of $q=0.23$, and a thin ($q=0.14$), flat
disk with a steep fall-off ($n=0.19$) in the profile beyond $r_e$.  Together
these two components add up to a total apparent magnitude of $m_{\rm
F547M}=13.1$, or roughly 10\% of the entire $V$-band output of the galaxy.  In
the disk plane, about $5\arcsec$ on both sides from the center, there appears
to be over-subtractions perpendicular to the major axis of the galaxy.
Because they are not fully symmetric across the galaxy plane nor across the
semi-minor axis, its suggests that there is another set of dust lanes that
gives rise to the pinched appearance in the disk in Figure~6{\it b}.  Our
decomposition shows that the bulge does not have a spherical component with a
de~Vaucouleurs profile, as suggested by Tsikoudi (1980), but rather is a
complicated blend of several disky and boxy components.

\subsubsection {NGC 4621}

NGC 4621 is an E5 galaxy with a very steep nucleus where $\left<\gamma\right> =
0.8$.  The nuclear cusp slope is defined by $\left<\gamma\right> \equiv
\mbox{dlog} (I)/ \mbox{dlog} (r)$ within $r < 0\farcs1$.  The isophotes are
disky but become gradually less so at large radii; the axis ratio decreases
outward from 0.88 to 0.64 in the same regime, and the PA remains essentially
constant with radius.  There are a number of globular clusters in the WFPC2
F555W image, but there is no evidence for nuclear dust.  Mizuno \& Oikawa
(1996) perform 2-D decomposition of a ground-based image of this galaxy and
find that it is made up of three distinct photometric components: an
exponential disk, a de~Vaucouleurs bulge, and an outer envelope.  The
exponential disk and de~Vaucouleurs bulge, which account for $16\%$ and $62\%$
of the total light, respectively, are both disky in shape, while the outer
envelope is spherical and contains $22\%$ of the galaxy luminosity.  Scorza \&
Bender (1995) and Simien \& Michard (1990) use two-component models, but they
arrive at different conclusions regarding the galaxy structure.  Simien \&
Michard (1990) find the disk scale-length to be $28\arcsec$, while Scorza \&
Bender (1995) obtain $7\farcs5$.  The discrepancy between the two studies may
be due in part to the use of different size images in the decomposition.

Figure~7{\it b}\ illustrates a fit using a Nuker bulge and an exponential disk
($r_s=4\farcs7$).  The fit produces residuals that have high symmetry.  Their
diffuseness and symmetry, suggest there is an intrinsic profile and shape
mismatch rather than there being additional components, in contrast to the
sharp features seen in Fig.~6{\it b} of NGC 4111.

Despite the changing galaxy shape and inflections in the surface brightness
profile, the entire image of NGC 4621 on the PC chip (inner $25\arcsec$
radius) can be subtracted remarkably well by employing two S\'ersic components
and a Nuker component (Fig.~7{\it c}).  The two innermost components have
similar scale-lengths, even though one is disky ($c=-0.35$) and steep
($n=5.98$) at the center and the other is boxy ($c=0.78$) and flat ($n=0.89$).
Both are required to fit the ``eye-shaped'' nucleus well, but our combination
may not be unique.  Hence, it is possible that the two may add to be a single
photometric component.  The outer envelope is characterized by a disky
($c=-0.16$) Nuker profile with $q=0.70$.  This component is not the spherical
halo component found by Mizuno \& Oikawa (1996) because our FOV is much
smaller.  In conjunction with their finding of a round galaxy halo, we support
the notion that NGC 4621 has $\sim$3, possibly 4, photometric components.

Upon casual visual inspection, the nucleus of NGC 4621 may appear to be 
pointlike because the isophotes are round and the profile is steep 
($\left<\gamma\right> = 0.8$).  But the three-component model fits the nucleus 
cleanly without the need to add in an unresolved point source.

\subsection {A ``Boxy'' Galaxy, NGC 5982}

The nuclear region (within $30\arcsec$) of this E3 galaxy has boxy isophotes
that become increasingly round at $r \le 3\arcsec$, but is otherwise uniform
with no sign of dust lanes (Figure~8).  The kinematics of this galaxy have
been studied in detail because the stellar velocities reverse direction
$2\arcsec-3\arcsec$ from the center, appearing as though the galaxy has a
kinematically distinct core with minor-axis rotation (Wagner, Bender,
M\"{o}llenhoff 1988).  One scenario to explain kinematically distinct cores
invokes mergers of galaxies that preserve their initial angular momentum
vectors.  An alternative is proposed by Statler (1991), who shows that
dynamically smooth, circulating orbits in a triaxial potential can, in
projection, cause the appearance of kinematic twists such as those observed in
NGC 5982.  To distinguish between the orbital projection model from the merger
model, Statler (1991) predicts that the former should produce a minimum in the
velocity field along a direction somewhere between the minor and major axis.
Moreover, there should be a photometric core that is similar in size to the
kinematic core.  Observations by Oosterloo, Balcells, \& Carter (1994) indeed
suggest that there is such a trough present in the velocity field at PA =
$78^\circ$.  But the ground-based surface photometry of this galaxy published
by Michard \& Simien (1988) and Bender, D\"{o}bereiner, \& M\"{o}llenhoff
(1988) do not reveal a photometric core down to at least $3\arcsec$.  Although
the kinematic peculiarity may be due to projection, Oosterloo et al. (1994)
suggest that the two effects: orbital projection and the merger of a small
clump, may not be mutually exclusive events.

With {\it HST}\ images, we find that the radial profile nearly flattens off
into a core interior to $0\farcs6$ (Fig.~8).  There also appears to be a
subtle break at $r\approx 3\arcsec$, where the nucleus becomes significantly
rounder.  The 3\arcsec\ radius also corresponds closely to the peak of the
rotation curve.  The finding that the photometric core is at least an order of
magnitude smaller than 7\arcsec\ rules out Statler (1991) models (T.~S.
Statler, 2002, private communication).  However, while the kinematically
decoupled core may not be best explained by those models, the orbital
projection effects proposed by Statler (1991) may still play a role.

Fig.~8{\it b}\ shows a one-component Nuker fit that produces large quadrupole
residuals at the center, but a relatively good fit farther out.  The
decomposition of this galaxy is complex and requires four components because
of the large ellipticity gradient that tends toward zero near the center.
Nonetheless, the galaxy can be accurately decomposed (Fig.~8{\it c}); the
slight quadrupole residual pattern (with RMS fluctuations 1\%-3\% of the
galaxy peak).  The ellipticity gradient in Figure~8 might be due to
triaxiality.  However, purely triaxial galaxies should also have confocal
isophotes, but we find that the first component is significantly displaced
from the rest by ($\Delta$RA, $\Delta$DEC) = ($0\farcs38,0\farcs16$),
suggesting there may have been a merger event proposed by Oosterloo et al.
(1994).  Forcing all the components to have the same position produces
asymmetric residuals near the center, and increases $\chi^2_{\nu}$ by
$3\times10^{-3}$ with $N_{\rm dof}=5.2\times10^5$.

About $13\farcs9$ (2.6 kpc in projected separation) from the center of NGC
5982, there is a galaxy that appears to have a double nucleus, although there
is a remote chance it may be caused by dust.  The dwarf galaxy's orientation
points nearly directly toward the center of NGC 5982.  If this galaxy is
indeed within the halo of NGC 5982, not just due to projection, one is tempted
to speculate that the dwarf companion is in the process of being tidally
disrupted by the larger galaxy.

To summarize, the profile of NGC 5982 flattens out into a core at
$\approx0\farcs6$, which is $\sim10$ times smaller than that predicted by
Statler (1991).  Although this finding rules out Statler (1991) models,
similar projection effects may still play a role in explaining the
kinematically decoupled core.  Our finding of an off-centered, luminous,
component at the core is probably evidence that this galaxy has had a recent
merger episode.

\subsection {Spiral Galaxies}

While early-type galaxies can be modeled well by ellipsoids, the obvious 
complication when fitting spiral galaxies is the presence of spiral arms and 
large-scale bars that are embedded in at least 50\%--60\% of the objects
(e.g., Eskridge et al. 2000).  Spiral arms cannot be easily parameterized with 
our models, but often there is a diffuse disk of older stars underneath the 
arms that can be well described by an exponential disk (e.g.,  de~Jong 
\& van~der~Kruit 1994; Prieto et al. 2001), whose contrast with the arms is
smaller in the near-infrared.  The situation is different for galactic bars,
as they can be fitted with a superposition of one or more analytic functions.
We show a decomposition of an ordinary spiral and a barred galaxy.

\subsubsection {NGC 4450}

NGC 4450 (Fig.~9) is an Sab galaxy viewed at a low inclination angle; its 
LINER nucleus exhibits broad emission lines (Stauffer 1982; Ho et al. 1997b), 
which appear double-peaked in {\it HST}\ spectra (Ho et al. 2000).  We will 
later revisit NGC 4450 in the context of point-source extraction (\S~5.2.3), 
but for now we focus only on the global galaxy morphology.  The 1-D 
decomposition by Baggett et al. (1998), assuming a spherical bulge and a 
circular disk, gives a B/D ratio of $\sim 0.3$.

The large scale disk of the galaxy extends well beyond the PC chip, thus we
first fit the entire WFPC2 mosaic image to get a better handle on the disk
parameters.  To do so, we mask out the echelon region missing in the WFPC2
FOV, and fit the remaining three quadrants plus the PC chip by assuming the
components are individually axisymmetric.  Then, holding the parameters of the
disk fixed, we fit only the PC chip while allowing the nuclear components to
optimize.  Decomposing the galaxy using the classic two-component model
(de~Vaucouleurs and exponential disk) gives B/D = 0.23, with $r_e=24\farcs0$
and $r_s=72\farcs2$, but the fit is poor.  In comparison to Baggett et al.
(1998), our scale-length measurement is about twice as large.  Part of this
large difference may be attributed to our inability to determine an accurate
background sky value because of the limited FOV of WFPC2; we set the sky level
to 26 mag arcsec$^{-2}$ (0.07 ADU).

An initial two-component decomposition of the WFPC2 mosaic reveals what
appears to be a bar.  This identification is reinforced in a three-component
fit (S\'ersic bulge, exponential disk, and a bar) that shows the bar to have a
flat profile within $r_e$ but steep fall off beyond it ($n=0.11$).  However,
the identification may not be secure because of the missing quadrant in the
mosaic.  For the purpose of decomposing the PC chip, this is not crucial, so
we assume that such a component exists.  After optimizing the parameters for
the disk and bar, we hold them fixed when fitting the bulge on the PC chip.
Through trial and error, we find that the bulge itself is better fit by
summing two S\'ersic components; neither one alone should be regarded with
high significance because the bulge is severely affected by dust.  If one were
to fit the bulge with a single component, Figure~9{\it b}\ shows that the
northern and southern regions of the nucleus would be oversubtracted.  The
central dark patches are probably oversubtractions, rather than dust
obscuration, because the round and bipolar shape of the residuals contrasts
sharply in appearance with the irregularity of the surrounding dust lanes.
With some confidence in the global components, we fit the remaining point
source on top of the nucleus, and the result of the fit is listed in Table 3.

To summarize, we use five components for NGC 4450: a bar, a disk, a nuclear 
point source, and a bulge made up of two S\'ersic functions.  In this 
decomposition, the bulge is somewhat fainter than the standard B/D 
analysis, in part because of the contribution of the bar, but also
because the nucleus is steeper than a de~Vaucouleurs profile.  We find 
B/D $\approx$ 0.09 if the bar is left out, or B/D $\approx$ 0.13 if it is 
included with the bulge.

\subsubsection {A Barred Galaxy, NGC 7421}

NGC 7421 is an SBbc galaxy seen at low inclination.  The bulge and the bar
are displaced from the center of the disky envelope, while the western bar
truncates in a ``bow-shock''-like spiral arm.  H~{\sc I} column-density maps
show a ``wake'' that suggests that NGC 7421 is plowing through, and interacting
with, the intracluster medium in which it is embedded (Ryder et al. 2000).

Decomposition of this galaxy in 2-D requires at least four components in order
to fit a bulge, a disk, and a bar (Fig.~10).  As with other spiral galaxies,
obtaining a plausible disk contribution requires using a FOV beyond that of
the PC.  So we proceed in a similar fashion to NGC 4450, by first fitting the
mosaic image, followed by optimizing over the PC region separately while
holding the large-scale disk component fixed.  A sky pedestal of 21.6 mag
arcsec$^{-2}$ (1.1 ADU) was determined from the edges of the mosaic.  Removing
the model of the galaxy reveals a faint dust lane that runs along the bar
through the nucleus.  Moreover, there is an unresolved compact nuclear
component that can be characterized by a Gaussian; this component has also
been noted by Carollo et al. (1998).  The ``bulge'' is composed both of an
intrinsic galaxy bulge plus the round component of the bar (components 2 and
3); the two are difficult to distinguish.  Without the fainter of the two
bulge components, Figure~10{\it b} shows doughnut-shaped residuals.  The
luminosity of the summed components of the ``bulge'' rivals that of the bar
(component 2), while the single most luminous component is the exponential
disk.  The bulge-to-disk ratio, excluding the bar component, is B/D $\approx$
0.06; including the bar with the bulge, B/D $\approx$ 0.15.

\section {EXTRACTION OF NUCLEAR POINT SOURCES}

\subsection {General Considerations}

A common application of galaxy decomposition is to extract reliable
measurements of nuclear point sources embedded in a host galaxy.  While
luminous quasars can overpower the light of the host galaxy, the opposite
extreme is the case for the more garden-variety AGNs in nearby galaxies, where
the nucleus is overwhelmed by the glare of the galaxy bulge.  Photometric
decomposition is absolutely crucial to properly disentangle the nuclear
component.  To do so, a central compact source or point source can be added on
top of the normal galaxy profile to represent the AGN (e.g., Kotilainen et
al.\ 1992; Portal 1999).  Implicit in this technique is the assumption that
the galaxy profile can be extrapolated all the way into the center by fitting
the asymptotic galaxy wings using standard functions such as the
de~Vaucouleurs plus exponential (Portal 1999) or Gaussian profiles (Carollo et
al.\ 1997).

When fitting faint nuclear point sources, the adopted form of the galaxy
profile is crucial because all functions differ widely in steepness near
$r=0$.  This is compounded by the fact it is difficult to fit profiles
accurately on large and small scales simultaneously; most of the weight is
dominated by outer regions where there are more pixels, at the expense of few
pixels in the center.  To overcome these complications, it is tempting to
simply exclude the outer regions.  But in so doing the intrinsic steepness of
the galaxy profile is less well constrained.  As galaxy nuclei continue to
rise as a power law even in the absence of central sources (Lauer et al.
1995), this practice easily produces fake compact source detections -- a power
law profile can always be artificially decomposed into a steeply rising line
plus a compact source.  Finally, if even the slowly varying region surrounding
the bulge cannot be well fit, the harder task of fitting the steeply rising
inner region should be even more dubious.  Therefore, the desired goal should
be to fit the entire nuclear region as accurately as possible when trying to
interpret and extract faint compact sources at the central few pixels.

We first compare the merits of 1-D and 2-D point-source extractions.  Some
arguments made in favor of 2-D techniques in the context of B/D decomposition
are directly applicable here.  The 1-D profile extracted from images is
affected by several factors: the seeing, the different ellipticities,
profiles, and shapes of the various galaxy components, a possible point
source, and other nuclear features such as dust.  There are few pixels near
the nucleus, so the inner isophotes are sensitive to small-scale features such
as dust, which can affect the centroiding and the shape of the nuclear
profile.  Miscentering can lead to uncertainties in the central source profile
in images where the PSF may be undersampled, as is the case with WFPC2.  As an
example of the ambiguity that can be encountered in extracting a weak nuclear
point source in 1-D, Figure~11 shows fits of the azimuthally averaged profiles
(without deconvolution) of NGC 4143 and NGC 4450.  Different assumptions can
also lead to conflicting evidence for a compact source.  In 1-D, for example,
it is impossible to tell whether the inflection at $r\approx 0\farcs2$ is
caused by circumnuclear dust or is a true sign of a photometrically distinct
compact component.  It is far easier to gauge the reality of compact sources
in 2-D based on the residuals of the model-subtracted image.

To compare the relative effectiveness of 1-D versus 2-D recovery techniques,
we added point sources to three real galaxies used in this study.  A point
source has only three free parameters ($x$ and $y$ position and brightness),
thus is easier to simulate and recover uniquely than compact sources,
providing the PSF is accurate.  Our simulations give only qualitative
guidelines for comparing the relative merits of 1-D and 2-D decompositions; we
do not cover all possible scenarios.  Other simulations of point-source
recovery have been done by Wadadekar et al. (1999), who use idealized galaxies
with exponential disks and S\'ersic profiles to explore a much wider range of
parameter space.

We simulate faint nuclear sources by placing a PSF on top of NGC 221, NGC
4621, and NGC 5982.  Figures~5{\it a}, 7{\it a}, and 8{\it a}\ show the
original 2-D images of the galaxies with contours overlaid; Figures 5{\it c},
7{\it c}, and 8{\it c}\ give the 2-D model residuals.  As shown in
Fig.~8{\it d} NGC 5982 has a global axis ratio of $q$ = 0.68, a slightly boxy
isophote shape, and a flat nuclear cusp of $\left<\gamma\right>=0.04$.  For
NGC 221, $q=0.72$ and $\left<\gamma\right>=0.5$.  For NGC 4621, $q=0.70$ at
large radius and is disky at $r\approx 8\arcsec$; the nucleus is rounded by
the seeing ($q=0.9$) and has a steep power-law cusp of
$\left<\gamma\right>=0.8$.  These examples have no apparent compact source at
the nucleus, no dust features, and span a wide range of nuclear slopes.  NGC
4621 appears to have a pointlike nucleus only because it has a very steep cusp
and because the nucleus is round.  We have shown in \S~4.3.2 that NGC 4621 can
be fit accurately without requiring a nuclear point source.  We can model all
three galaxies quite well in both 1-D (using a single Nuker profile) and 2-D.

We add Tiny Tim PSFs with brightnesses 2 to 7 magnitudes fainter than the
bulge, in 1 magnitude intervals, then extract them with the same PSF.  The
bulge magnitude is defined here as the flux within $r\le0\farcs5$.
Figure~12{\it a}--15{\it c}\ illustrate the projected profile resulting from
adding the point sources.  Figure~13 compares the results of the extractions
in 1-D versus 2-D, showing deviations between input and recovered magnitudes.
The S/N of the images are as high as $\sim 70-80$ at the peak.  When the point
source is bright and the nucleus is moderately flat, the 1-D technique can
recover the point sources well.  But the recovery in 1-D becomes poor very
quickly for a moderately dim source and a steep profile.  For
$\left<\gamma\right>\gtrsim0.5$ there is little accuracy even for relatively
luminous sources.  At the steep end of $\left<\gamma\right>$, the technique is
completely unreliable.

By comparison, 2-D modeling can recover much fainter nuclear
sources, although there is a slight systematic offset.  The recovered
magnitude in most cases is slightly fainter than the input magnitude, a
consequence of the degeneracy between a steep galaxy profile (parameterized
using a Nuker function) and a point source.  The use of the Nuker function can
overfit a steep cusp at the expense of a point source.  For faint point
sources and steep cusps, the opposite scenario may occur; the recovered
magnitude may be brighter than the input at the expense of a shallower cusp.

In AGN studies, it is often informative to know the spatial extent of the
nuclear source.  Direct non-stellar emission from a central engine is expected
to be unresolved, even for the nearest AGNs, whereas reprocessed non-stellar
light or a nuclear star cluster may be resolved.  To test the source extent,
one can fit an analytical PSF function, such as the Moffat profile described
in \S~2.3.2, with an adjustable width.  In our simulations in 1-D, we find
that the luminosity and width of the compact source depend sensitively on
initial parameters because of degeneracies in the fits.  We conclude that 1-D
fitting generally gives unreliable measures of compactness for central
sources.  In our 2-D simulations, the widths converge to that of a point
source in every case, except for the faintest test sources.

For the purpose of extracting faint compact sources, we can constrain their
parameters by exploring the degeneracies of the local minima in $\chi^2$, and
by fitting different nuclear models.  In fact, the latter is a necessary
exercise because different model assumptions may produce contradictory results
because of profile ambiguities.  For instance, a compact source on top of an
exponential bulge may, in some cases, be mimicked by a single Nuker profile
without a compact source.  In the next section we give some examples of 
the 2-D technique of fitting point sources.

\subsection {Examples}

\subsubsection {NGC 2787} 

The SB0 galaxy NGC 2787 contains a LINER nucleus with broad Balmer lines (Ho
et al. 1997b).  The association of this object with AGN activity is
corroborated by the recent {\it HST}\ detection of a central dark mass of
$7\times10^{7}M_\odot$, presumed to be a massive black hole (Sarzi et al.
2001).  The smooth and otherwise featureless bulge is beautifully adorned by a
circumnuclear dust disk with ripples that lie nearly orthogonal to the major
axis of the bulge (Fig.~14).  Isophotal analysis shows a dip between $1\arcsec
- 6\arcsec$ in the profile cut and shape parameters, caused by the dust disk.
The bulge itself has an ellipticity of 0.32 at large radii with a large and
fluctuating gradient; the gradient is due in part to dust.

In 2-D, the bulge can best be fit by an exponential and a Nuker component.
Although the bulge component can also be well fitted by a superposition of a
S\'ersic ($m_{\rm F547M}=12.43$ mag, $n=2.71$, and $r_e=7\farcs43$) and an
exponential ($m_{\rm F547m}=11.56$ mag, $r_s=11\farcs74$), overall the first
combination gives formally a better fit.  We list their parameters in Table 3.
Figure~14{\it b}\ shows a fit where we use a single Nuker component.  Although
most of the bulge is removed, considerable over- and under-subtractions
remain.

Various trials suggest that there is a compact source at the center of the
nucleus, but there is considerable ambiguity as to whether it is resolved.
While we can force the object to be a true point source in the fit, there is a
small but significant pedestal in the residuals.  The compact source is
resolved in an unconstrained fit.  However, we cannot rule it out as being
unresolved because of its faintness and because the excess pedestal may be
caused by model mismatch near the galaxy core. If a point source is present,
we can limit its magnitude to $m_{\rm F547M}\gtrsim19.8$ mag, or $M_V\gtrsim
-10.8$ mag.

Another interesting component that remains in the residuals is a luminous 
stellar disk at the center of the dust disk.  The stellar disk is flat and has 
sharp, truncated edges of 100 pc ($1\farcs6$) in semi-major axis.  The 
sharpness of the disk truncation, the uniformity of the disk, and the clean 
subtraction around its immediate surroundings all suggest the disk is likely 
to be real, rather than an image artifact created by a model mismatch.

\subsubsection{NGC 4450}

The overall fit of NGC 4450 was discussed in \S~4.5.2.  Surface photometry of
this object (Fig.~9{\it d}) shows a double power law with a break at
$r\approx 0\farcs7$.  The apparent power-law behavior of the inner region is
only slightly violated by an inflection at $r=0\farcs2$, which hints at the
presence of a compact central source.  However, Figure~9{\it c}\ shows that
the inflection in the 1-D profile may be caused by the extensive dust present
at small radii.  The situation is much clearer in 2-D.  Our modeling finds a
nuclear point source with $m_V=18.9\pm0.2$ mag, or $M_V=-12.2\pm0.2$ mag.

\section {SUMMARY}

We have presented a general algorithm to decompose a galaxy into components in
two dimensions. To illustrate its merits and flexibility, we have applied it
to high-resolution optical {\it HST}\ images of 11 galaxies of various shapes
and morphological types.  As another example, Peng (2002) presents a detailed
decomposition of the double nucleus of M31.  We show that we can model the
central regions of galaxies accurately, even for difficult cases that show
large isophotal twists and changes in shape.  These morphological complexities
are often signatures of distinct galaxy components.  After the major
components are removed, in a number of systems we discover evidence of galaxy
substructures too subtle to be seen in the original image. These include
features such as nuclear point sources, low-level dust patterns, stellar
disks, stellar bars, and other distinctive large-scale components.  Despite
the featureless appearance of some giant elliptical galaxies (e.g., NGC 4621,
NGC 5982), in detail they reveal unusual shapes and slight misalignment of the
sub-components.  The physical interpretation of these subtle features is not
yet clear.  In theory, the amount of distortion in a galaxy --- as evidenced,
for example, by the need for fitting multiple components, their different
shapes and displacements --- may give clues to its evolutionary history.
Substructure may also signal triaxiality in the bulge potential.  A fruitful
avenue of future research would be to compare the structural decomposition of
real galaxies with a similar analysis applied to high-resolution N-body
simulations of galaxy formation.  Another is to couple the 2-D structural
analysis with integral-field kinematic maps.

\acknowledgments

We thank Alice Quillen for enlightening discussions about various galaxies in
our sample, and Dennis Zaritsky, Daniel Eisenstein, and Adam Burrows for
discussions about error analysis. We also thank the referee for comments to
improve this work.  L.~C.~H. acknowledges financial support through NASA
grants from the Space Telescope Science Institute (operated by AURA, Inc.,
under NASA contract NAS5-26555).  This research has made use of the NASA/IPAC
Extragalactic Database (NED) which is operated by the Jet Propulsion
Laboratory, California Institute of Technology, under contract with the
National Aeronautics and Space Administration.

\newpage


\footnotesize
\begin{verbatim}
===============================================================================
# IMAGE PARAMETERS
A) f547m.fits          # Input data image (FITS file)
B) f547m-out.fits      # Name for the output image
C) none                # Noise image name (made from data if blank or "none") 
D) f547m-psf.fits      # Input PSF image for convolution (FITS file)
F) dust                # Pixel mask (ASCII file or FITS file with non-0 values)
G) 47 800 57 800       # Image region to fit (xmin xmax ymin ymax)
H) 427 435             # Convolution box center
I) 60   60             # Size of convolution box (x y)
J) 21.689              # Magnitude photometric zeropoint 
K) 0.046   0.046       # Plate scale (dx dy).  Relevant only for Nuker model.

# INITIAL FITTING PARAMETERS
#
#   For object type, allowed functions are: sersic, nuker,
#                       expdisk, devauc, moffat, gaussian.
#

# Objtype:      Fit?         Parameters

===============================================================================

 0) nuker              # Object type
 1) 427.17  435.46 0 1 # position x, y    
 3) 12.16      1       #    mu(Rb)
 4) 4.45       1       #     Rb
 5) 1.62       1       #    alpha 
 6) 1.26       1       #     beta 
 7) 0.21       1       #    gamma 
 8) 0.72       1       # axis ratio (b/a)  
 9) 39.48      1       # position angle (PA) 
10) 0.00       0       # diskiness/boxiness 
 Z) 0                  # Output image type (0 = residual, 1 = Don't subtract) 

 0) sersic             # Object type
 1) 427.17  435.46 1 1 # position x, y    
 3) 15.22      1       # total magnitude    
 4) 39.70      1       #     R_e            
 5) 3.03       1       # sersic index (deVauc=4)  
 8) 0.9        1       # axis ratio (b/a)   
 9) -18.75     1       # position angle (PA)
10) -1.70      1       # diskiness/boxiness
 Z) 0                  # Output image type (0 = residual, 1 = Don't subtract) 

 0) expdisk            # Object type
 1) 427.17  435.46 0 0 # position x, y    
 3) 14.96      1       # total magnitude
 4) 63.76      1       #     Rs
 8) 0.75       1       # axis ratio (b/a)   
 9) -34.03     1       # position angle (PA)
10) 0.30       1       # diskiness/boxiness
 Z) 0                  # Output image type (0 = residual, 1 = Don't subtract) 

 0) gaussian           # Object type 
 1) 427.17  435.46 1 1 # position x, y 
 3) 20.50      1       # magnitude          
 4) 0.50       0       #    FWHM             
 8) 1.0        1       # axis ratio (b/a)   
 9) -34.03     1       # position angle (PA)
10) 0.00       1       # diskiness/boxiness)
 Z) 0                  # Output image type (0 = residual, 1 = Don't subtract) 

 0) sky
 1) 0.60       1       # sky background 
 Z) 0                  # Output image type (0 = residual, 1 = Don't subtract) 

\end{verbatim}
\figcaption[]{Example of an input file for GALFIT.}
\normalsize


\vfill\eject

\begin{deluxetable}{ccccc}
\tablewidth{0pt}
\tablecaption {Computational Requirements for 2-D Galaxy Modeling\tablenotemark{a}}
\tablehead{ Fitting Region Size & Convolution Size & Memory & Total Time & 
Functions Used\\
(pix$\times$pix) & (pix$\times$pix) & (Mbytes) & (minutes) &  \\
(1) & (2) & (3)  & (4) & (5) }
\startdata 
$750\times750$ & $128\times128$ & 55 & 6--10 & S\'ersic, Expdisk, sky \\ 
$350\times350$ & $128\times128$ & 20 & 3 -- 5 & S\'ersic, Expdisk, sky \\ 
$200\times200$ & $128\times128$ & 14 & $<1$ -- 3 & S\'ersic, Expdisk, sky \\ 
\tableline
$750\times750$ & $256\times256$ & 60 & 10--15 & S\'ersic, Expdisk, sky \\ 
$350\times350$ & $256\times256$ & 30 & 6 -- 10 & S\'ersic, Expdisk, sky \\ 
$200\times200$ & $256\times256$ & 20 & 4 -- 7 & S\'ersic, Expdisk, sky \\ 
\tableline
$750\times750$ & $128\times128$ & 70 & 20--30 & Nuker, S\'ersic, S\'ersic, sky \\ 
$350\times350$ & $128\times128$ & 30 & 10--20 & Nuker, S\'ersic, S\'ersic, sky \\ 
$200\times200$ & $128\times128$ & 20 & 3--10 & Nuker, S\'ersic, sky \\
\tableline
\enddata 
\tablenotetext{a}{Estimated computation requirements on an Intel Pentium 450 MHz
computer.}
\tablecomments {
Col. (1): The size of the data image being fitted.  
Col. (2): Size of the convolution box, including all the necessary padding 
(see text).
Col. (3): The amount of memory used by GALFIT.  
Col. (4): Cumulative fitting time.  
Col. (5): Functions fitted simultaneously.
}
\end{deluxetable}


\begin{figure} 
    \plottwo {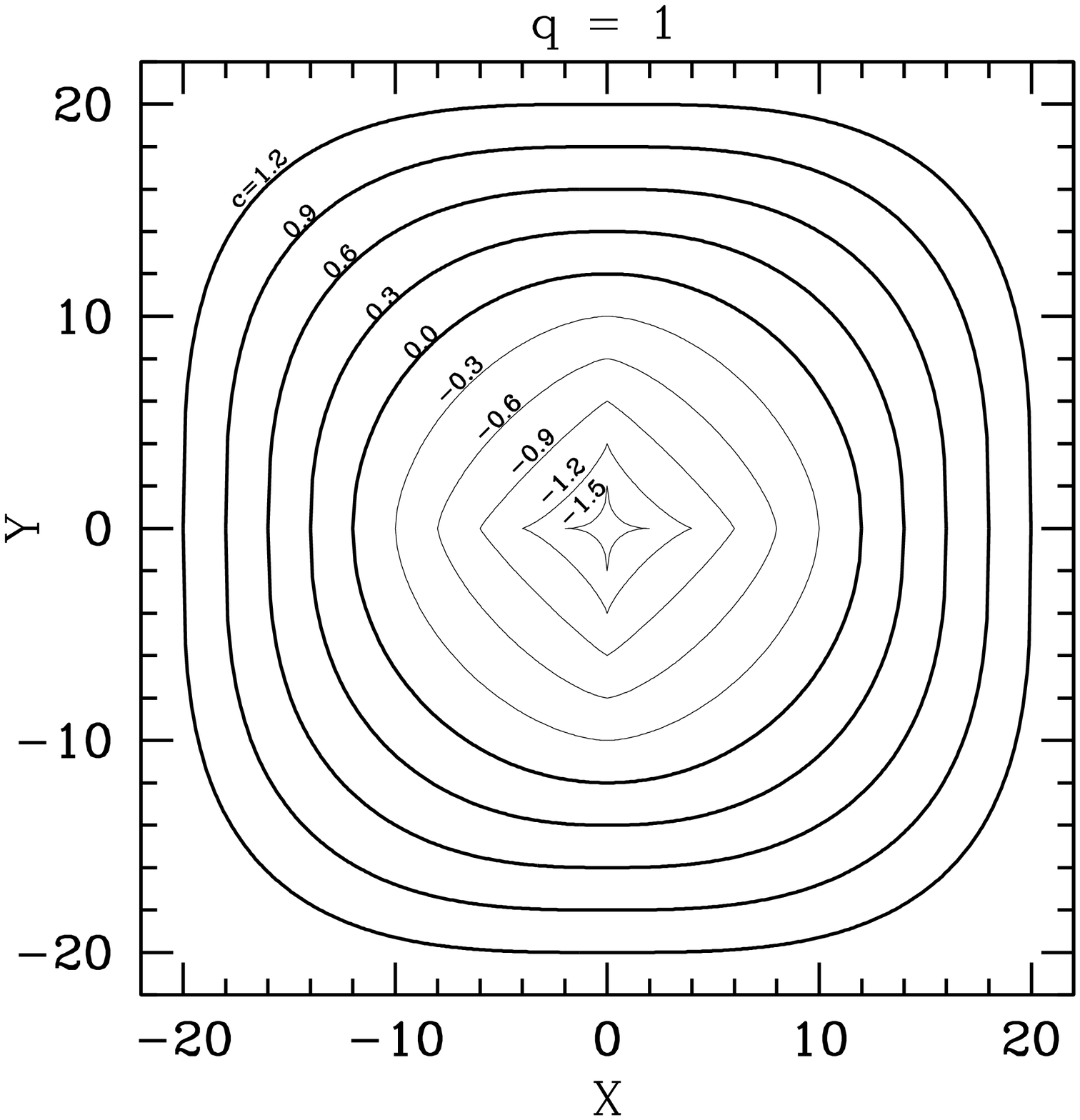}{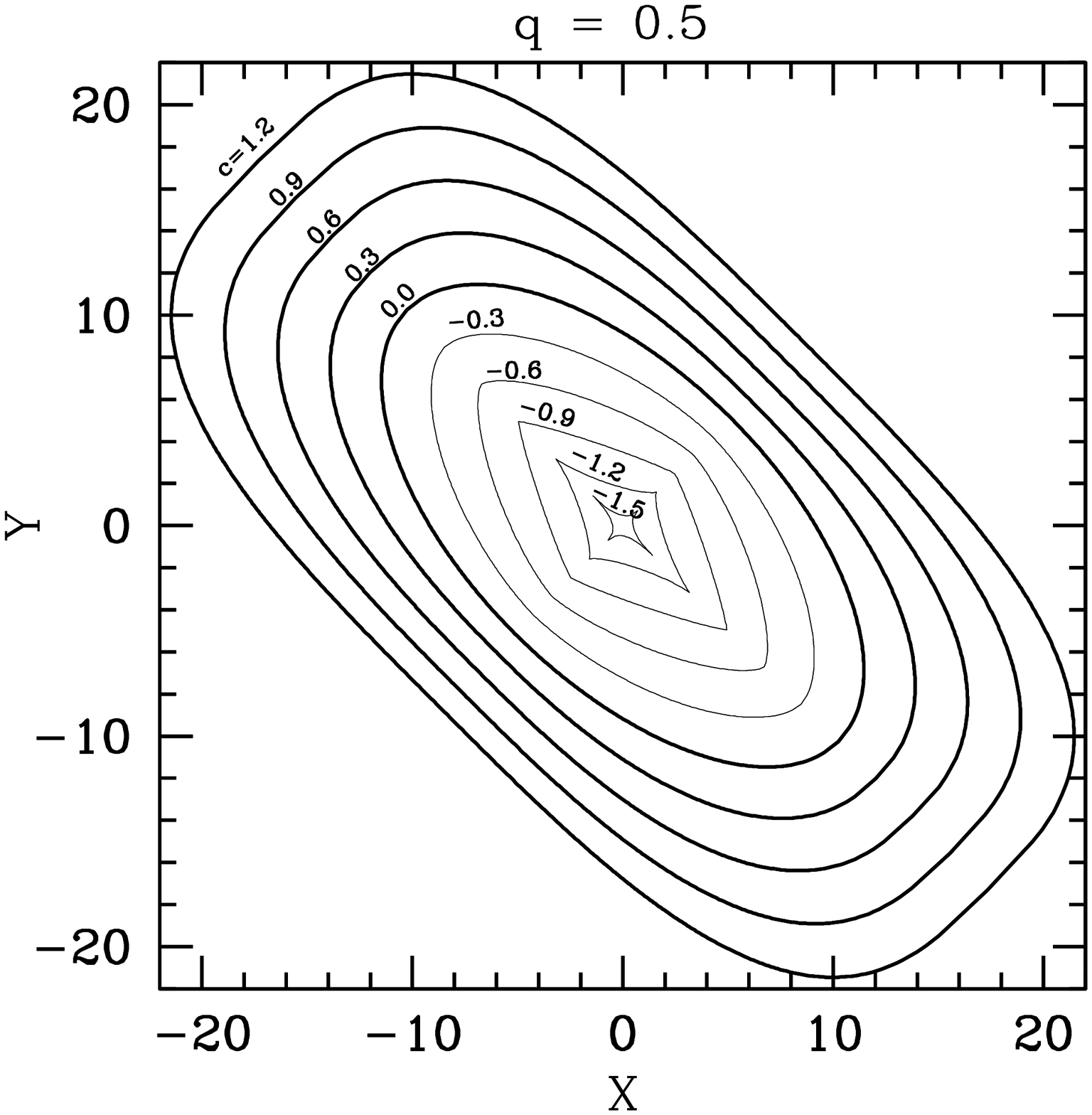} 
    \figcaption[cpeng-galfit.fig2.eps] {The azimuthal shape of ellipses based on
		   Equation 1 for two different axis ratios $q=1$ ({\it left}) 
                   and $q=0.5$ ({\it right}), as a function of the 
                   diskiness/boxiness parameter $c$}
\end{figure}


\begin{figure}
    \plottwo {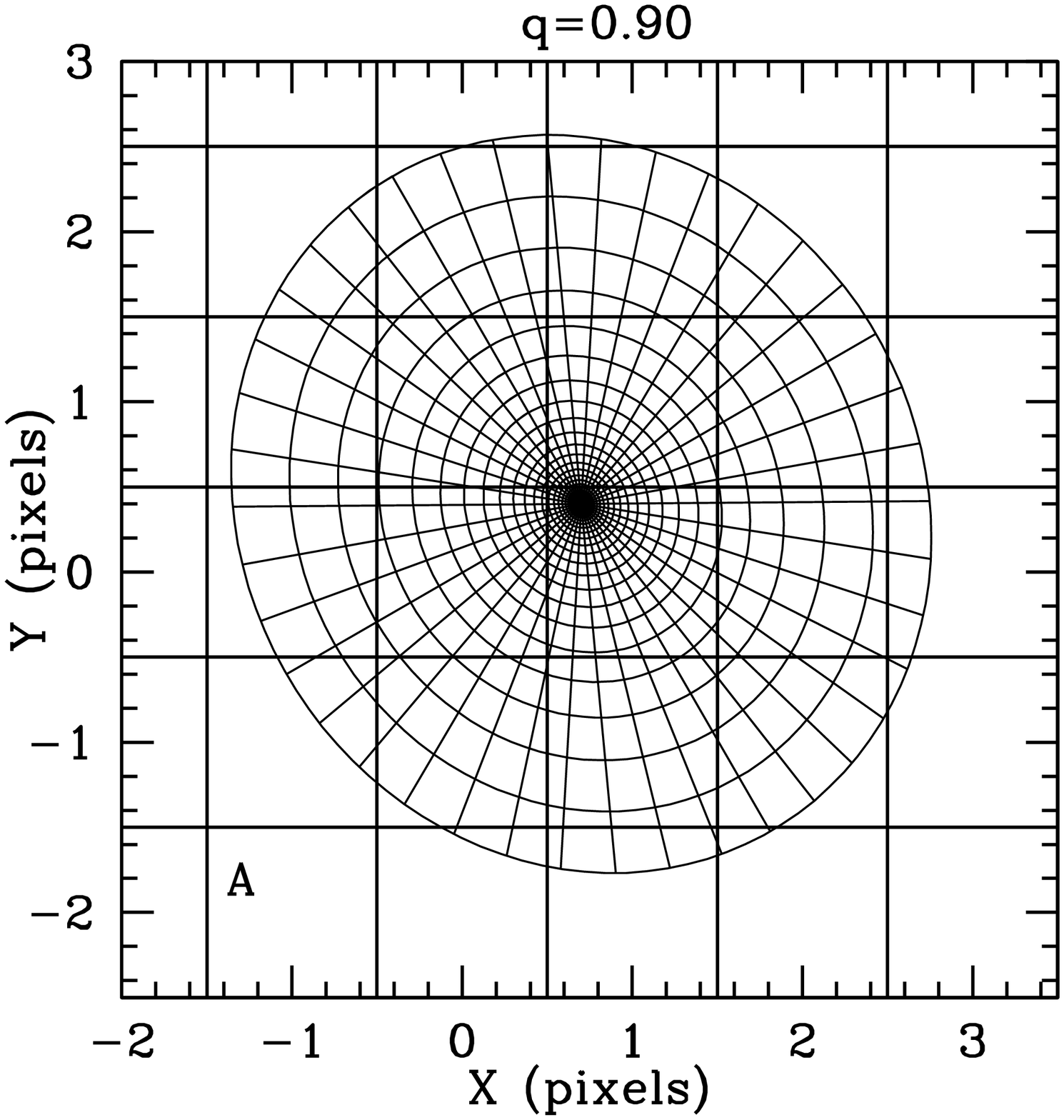}{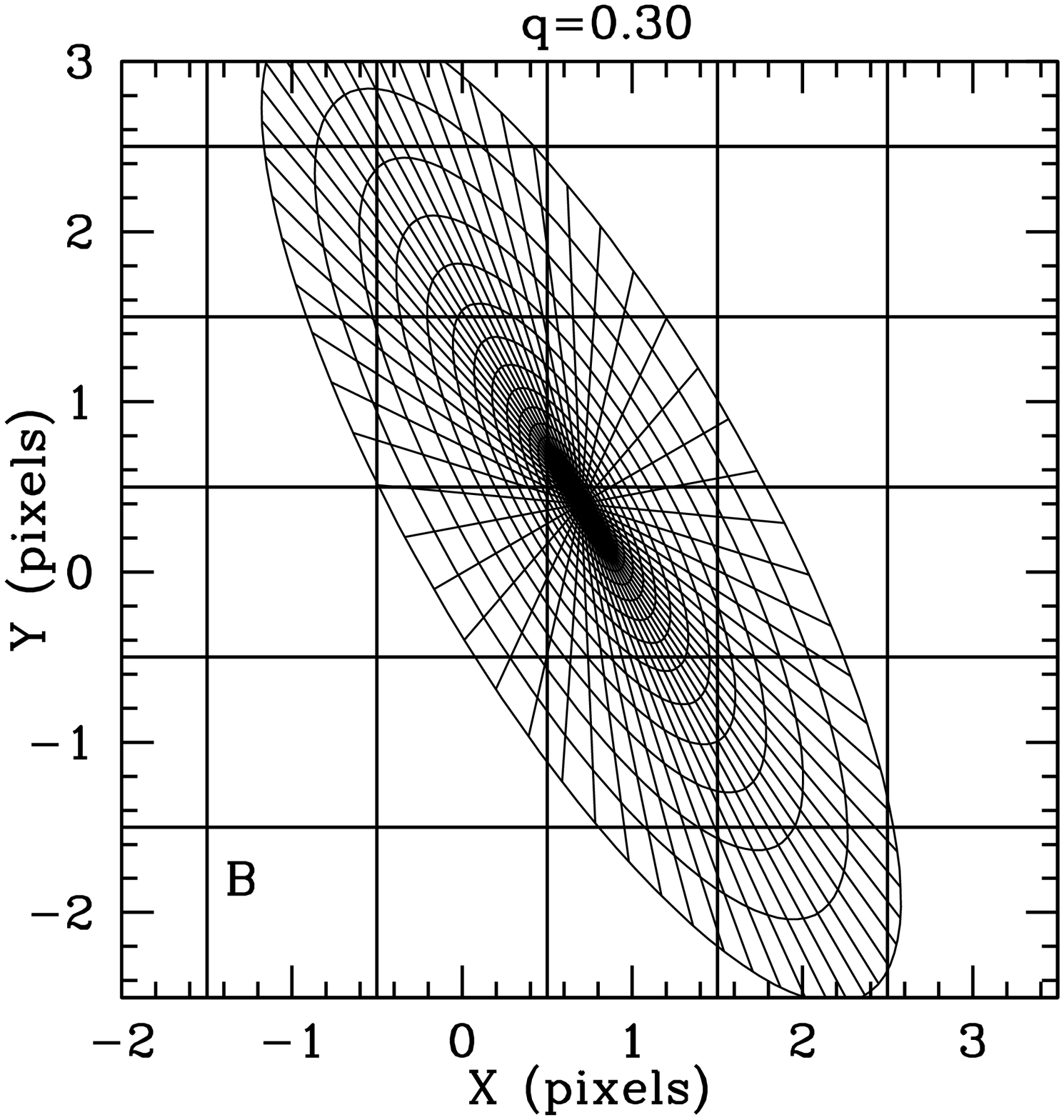}
    \figcaption[cpeng-galfit.fig3a.eps, cpeng-galfit.fig3b.eps] {Elliptical polar-gridding
		   method of integrating pixel values near $r=0$ for axis
		   ratio ({\it a}) $q=0.9$ and ({\it b}) $q=0.3$.  The angular
		   and radial spacing of the grids are arbitrary in these
		   pictures.}
\end{figure}

\begin{figure}
    \plotone{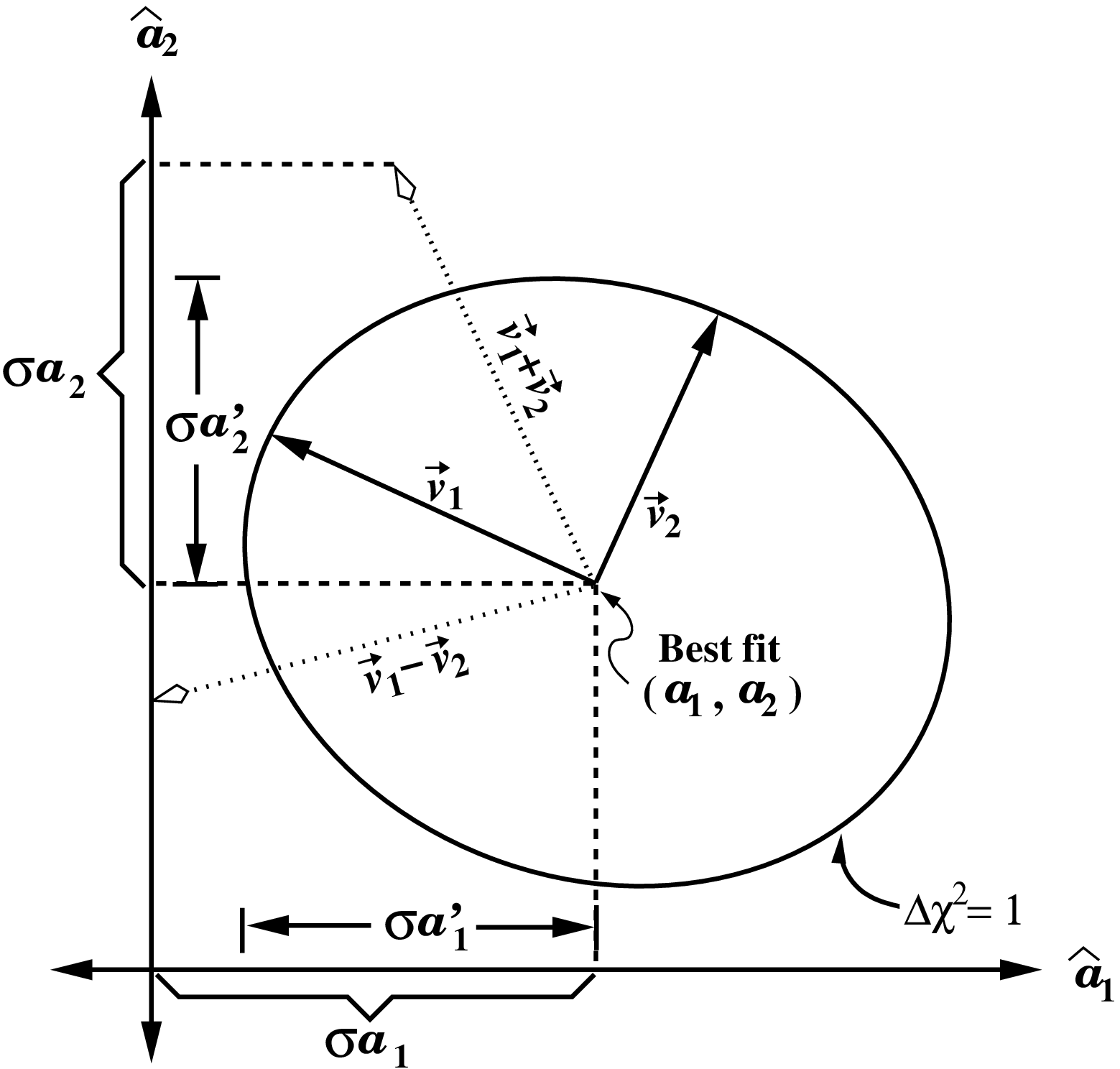} 
    \figcaption[cpeng-galfit.fig4.eps] {Estimating uncertainties for a fit of two 
                   correlated parameters, with best fit values $a_1$ and 
                   $a_2$.  The
		   $\Delta\chi^2=1$ ellipse has principle axes $\vec{v}_1$ and
		   $\vec{v}_2$, and bounds a 68\% confidence region on
		   individual parameters.  $\sigma a_1$ and $\sigma a_2$ are
		   the largest projected vector sums of $\vec{v}_1$ with
		   $\vec{v}_2$ onto $\hat{a}_1$ and $\hat{a}_2$, respectively;
		   they are the uncertainties we
		   quote for $a_1$ and $a_2$, even though the true
		   uncertainties are strictly speaking $\sigma a_1\arcmin$ and
		   $\sigma a_2\arcmin$ (see text).}
\end{figure}

\newpage

\begin{deluxetable}{lrcccr}
\tablewidth{0pt}
\tablecaption {Basic Galaxy Data}
\tablehead{Galaxy&$D$ & Hubble Type & Spectral Class & Filter & $t_{\rm exp}$ \\
    & (Mpc) & &     &     & (s) \\
(1) & (2) & (3)  & (4) & (5) & (6) }
\startdata
NGC 221 (M32) & 0.7     & E2   & A     & F555W & 104 \\
NGC 2787      & 13.0    & SB0+ & L1.9  & F547M & 360 \\
NGC 4111      & 17.0    & S0+  & L2    & F547M & 300 \\
NGC 4450      & 16.8    & Sab  & L1.9  & F555W & 520 \\
NGC 4621      & 16.8    & E5   & A     & F555W & 330 \\
NGC 5982      & 38.7    & E3   & L2    & F555W & 1000 \\
NGC 7421      & 24.4    & SBbc &       & F606W & 600 \\
\tableline
\enddata 

\tablecomments {
Col. (1): Galaxy name.
Col. (2): Adopted distance as given in Tully 1988 or otherwise derived from the 
heliocentric radial velocity and $H_o=75\, \mbox{km s}^{-1}\mbox{Mpc}^{-1}$. 
Col. (3): Hubble type from de~Vaucouleurs et al. 1991.
Col. (4): Spectral class of the nucleus from Ho et al. 1997a, where A = 
absorption-line nucleus and L = LINER.  Type~2 objects have no broad emission 
lines and ``type 1.9'' objects have weak broad H$\alpha$ emission.
Col. (5): {\it HST}\ filter.
Col. (6): Exposure time.
}
\end{deluxetable}
\vfill\eject


\begin{deluxetable}{llrrrrrrrrrrrrrrl}
\rotate
\tablewidth{0pt}
\tablecaption {Two-Dimensional Image Fitting Parameters}
\tablehead{ Galaxy & Filter & Func. & $f/f_{\rm tot}$ & $\Delta\alpha$ & $\Delta\delta$ & Mag & 
$r_{\hbox{b,e,s}}$ & $\alpha, n$ & $\beta$ & $\gamma$ & $q$ & PA & 
$c$ & $\chi^2_\nu$ & $\Delta\chi^2_\nu$ &Comments \\
 Mag ($r<10\arcsec$)  &   &   &  & ($\arcsec$) & ($\arcsec$) & & ($\arcsec$) & & & & & (deg) & & & & \\
(1)    & (2) & (3) & (4) & (5) & (6) & (7) & (8) & (9) & (10) & (11) & (12) & (13) & (14) & (15) & (16) & (17) }
\startdata
NGC 221  & F555W & S\'ersic & 0.94 & $\equiv 0.$ & $\equiv 0.$ &  7.51 & 138.97 & 6.69 &      &      & 0.72 &$-20.4$ &$-0.06$& 1.00 & 0.13 \\
  9.95   &       & Exp't    & 0.06 &       0.01  &      $-0.01$& 12.96 &  0.41  &      &      &      & 0.72 &$-21.2$ &$-0.05$& \\
\tableline
NGC 2787 & F547M & Exp't    & 0.38 & $\equiv 0.$ & $\equiv 0.$ & 12.56 &   7.11 &      &      &      & 0.60 &$-62.8$ & 0.32  & 0.81 & 0.01 \\
 12.24   &       & Nuker    & 0.68 &       0.56  &     $-0.33$ & 16.92 &   1.10 & 0.42 & 1.57 & 0.44 & 0.77 &$-67.1$ &$-0.16$& \\
         &       & Gauss    & 0.001&       0.58  &     $-0.35$ & 19.78 &   0.07 &      &      &      & 1.0  &  0.0   & 0.0   & \\
\tableline
NGC 4111 & F547M & Exp't    & 0.20 & $\equiv 0.$ & $\equiv 0.$ & 11.05 &  21.39 &      &      &      & 0.12 &$-31.6$ &$-0.65$& 1.73 & 0.31 \\
 11.46   &       & S\'ersic & 0.60 &        0.16 &     $-0.22$ & 11.58 &   7.85 & 2.51 &      &      & 0.56 &$-32.0$ & 0.28  & \\
         &       & S\'ersic & 0.18 &        0.16 &     $-0.15$ & 13.33 &   4.49 & 0.48 &      &      & 0.23 &$-31.1$ &$-0.58$& \\
         &       & S\'ersic & 0.36 &        0.17 &     $-0.18$ & 14.84 &   4.52 & 0.19 &      &      & 0.14 &$-31.7$ &$-0.75$& \\
         &       & S\'ersic & 0.003 &       0.17 &     $-0.14$ & 17.82 &   0.46 & 0.19 &      &      & 0.42 &$-21.6$ &$-1.68$& \\
\tableline
NGC 4450 & F555W & Exp't    & 0.15 & $\equiv 0.$ & $\equiv 0.$ &  9.44 &  93.55 &      &      &      & 0.57 & $-7.4$ &$-0.20$& 1.59 & 0.19 \\
 12.42   &       & S\'ersic & 0.74 &     $-0.10$ &       0.04  & 12.08 &  10.40 & 2.06 &      &      & 0.74 &   3.3  &$-0.01$&      &      & Bulge \\
         &       & S\'ersic & 0.07 &        0.02 &     $-0.01$ & 15.19 &   0.90 & 2.24 &      &      & 0.83 &  46.3  &  0.38 &      &      & Bulge \\
         &       & S\'ersic & 0.07 & $\equiv 0.$ & $\equiv 0.$ & 12.98 &  31.74 & 0.11 &      &      & 0.40 &  12.8  &  0.89 &      &      & Bar \\
         &       & PtSrc    & 0.003 &        0.0 &         0.0 & 18.86 &   0.01 &      &      &      & 1.0  &   0.0  &  0.00 & \\
         &       & Sky      &       &            &             & 26.00 &        &      &      &      &      &        &       &      &      & mag/arcsec$^2$ \\
\tableline
NGC 4621 & F555W & S\'ersic & 0.12 & $\equiv 0.$ & $\equiv 0.$ & 13.74 &   2.90 & 5.98 &      &      & 0.78 &$-17.5$ &$-0.35$& 1.02 & 0.02 \\
 11.66   &       & S\'ersic & 0.04 &        0.03 &     $-0.11$ & 15.50 &   3.57 & 0.89 &      &      & 0.20 &$-16.1$ & 0.78  & \\
         &       & Nuker    & 0.85 &     $-0.01$ &        0.03 & 18.84 &  12.13 & 1.10 & 1.87 & 0.70 & 0.70 &$-16.5$ &$-0.16$& \\
\tableline
NGC 5982 & F555W & S\'ersic & 0.15 & $\equiv 0.$ & $\equiv 0.$ & 13.46 &  13.21 & 0.67 &      &      & 0.94 &$-64.3$ &  0.67 & 1.27 & 0.82 \\
 12.45   &       & S\'ersic & 0.16 &        0.38 &     $-0.16$ & 14.48 &   3.52 & 0.58 &      &      & 0.91 &$-69.6$ &  0.62 & \\
         &       & S\'ersic & 0.06 &        0.38 &     $-0.18$ & 15.45 &   0.79 & 0.96 &      &      & 0.83 &  14.2  &  0.52 & \\
         &       & Nuker    & 0.63 &        0.37 &     $-0.17$ & 17.15 &   1.22 & 1.29 & 1.51 & 0.00 & 0.59 &$-72.6$ &  0.23 & \\
\tableline
NGC 7421 & F606W & Exp't    & 0.43 & $\equiv 0.$ & $\equiv 0.$ & 12.00 &  23.00 &      &      &      & 0.91 &$-82.3$ &  0.0  & 1.18 & 1.97 \\
 13.89   &       & S\'ersic & 0.23 &     $-3.65$ &     $-1.46$ & 14.63 &  20.34 & 0.40 &      &      & 0.21 &$-90.4$ &$-0.71$&      &      & Bar\\
         &       & S\'ersic & 0.25 &     $-2.57$ &     $-1.43$ & 15.42 &   3.14 & 0.74 &      &      & 0.73 &$-90.5$ &  0.32 &      &      & Bulge \\
         &       & S\'ersic & 0.08 &     $-2.34$ &     $-1.39$ & 16.57 &   1.05 & 2.90 &      &      & 0.92 &$-97.8$ &$-0.10$&      &      & Bulge\\
         &       & Gauss    & 0.030 &    $-2.34$ &     $-1.40$ & 20.07 &   0.02 &      &      &      & 1.0  &$-90.4$ &  0.0  &      &      &  \\
         &       & Sky      &      &             &             & 21.58 &        &      &      &      &      &        &       &      &      & mag/arcsec$^2$ \\
\tableline
Uncertainties &  &          &      & $\lesssim0.002$ & $\lesssim 0.002$ & $\lesssim 0.05$  & $\lesssim 1\%$  & $\lesssim 0.05$  & $\lesssim 0.05$ & $\lesssim 0.05$ & $\lesssim 0.01$ &  $\lesssim 0.2$ & $\lesssim 0.01$ &      &      & \\
\tableline
\enddata 
\tablecomments {
Col. (1): Galaxy name and total aperture magnitude within $r=10\arcsec$ of
          galaxy center.
Col. (2): {\it HST}\ filter. 
Col. (3): Galaxy components used in the fit.  
Col. (4): The ratio of component flux to the total flux within a $r=10\arcsec$
          aperture of the galaxy center.
Col. (5): R.A. offset.  
Col. (6): Dec. offset.
Col. (7): For Nuker, it is the surface brightness at the breaking radius.  For
	  all else, it is the total brightness of that component.  The
	  magnitudes are not corrected for Galactic extinction.
Col. (8): $r_{\hbox{b}}$ is the break radius for the Nuker power law;
	  $r_{\hbox{e}}$ is the effective radius of the S\'ersic law;
	  $r_{\hbox{s}}$ is the scale-length of the exponential disk; for a 
	  Gaussian it, is the FWHM.  
Col. (9): For Nuker, $\alpha$ parameterizes the sharpness of the break.  For
	  S\'ersic, $n$ is the S\'ersic exponent 1/$n$.
Col. (10): Asymptotic outer power-law slope of the Nuker law.  
Col. (11): Asymptotic inner power-law slope of the Nuker law.  
Col. (12): Axis ratio.  
Col. (13): Position angle.  
Col. (14): Diskiness (negative)/boxiness (positive) parameter.
Col. (15): Best $\chi^2_\nu$ for fits shown in image panels {\it c}.
Col. (16): $\chi^2_\nu$ increase from best fit, for fits shown in image panels {\it b}.
Col. (17): Comments.
           The last row gives a ``representative'' uncertainty estimate (as
	   described in \S~2.6) for the ensemble.  For most parameters, the
	   uncertainties cited are conservative estimates.
}
\end{deluxetable}

\vfill\eject


\begin{figure}
    \hbox{
        \hskip -0.4truein
        \includegraphics[scale=2.0,angle=90]{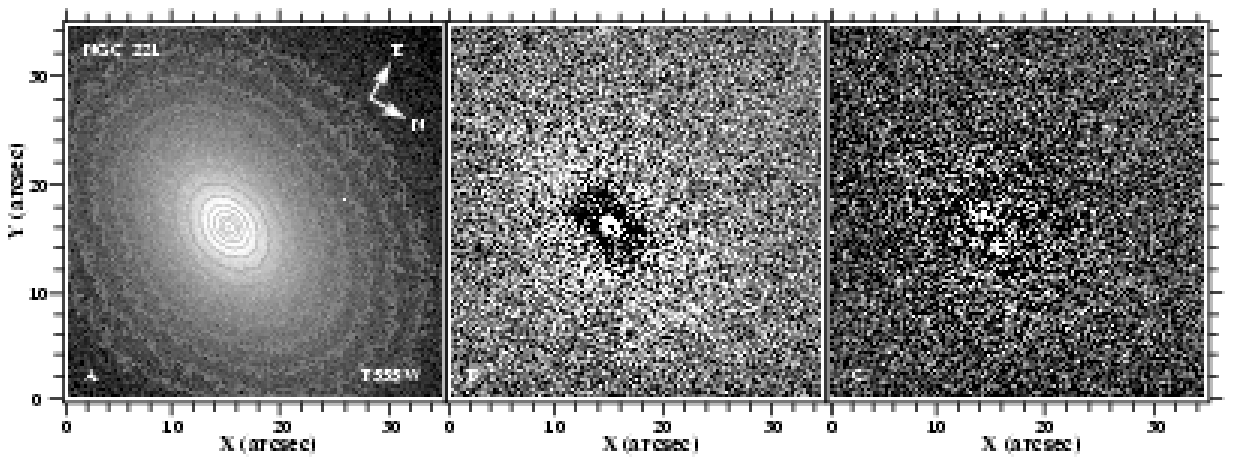} 
        \raisebox{1.5truein}{
        \includegraphics[width=4.0truein,height=7.truein]{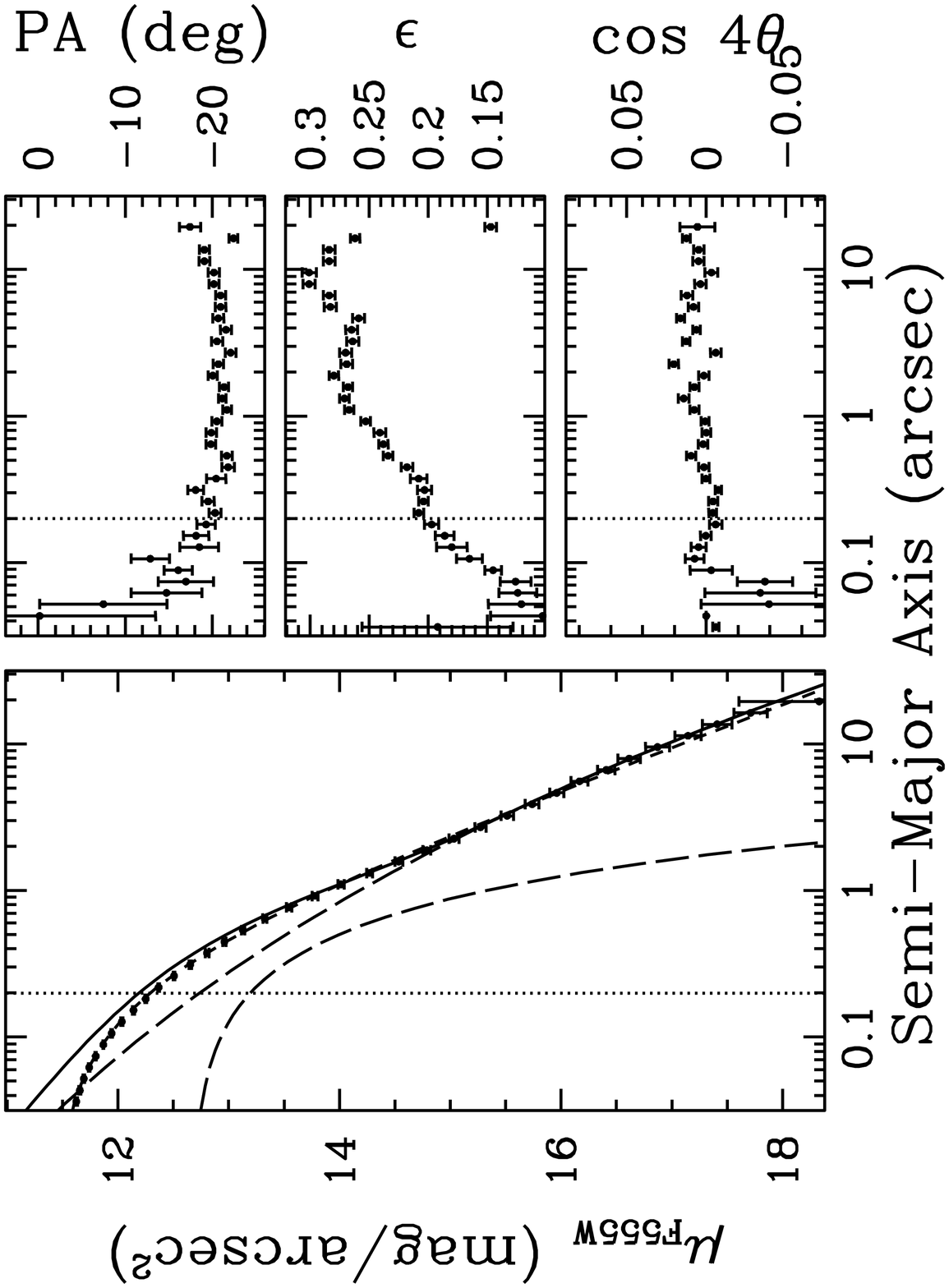}}
    } 

    \figcaption[cpeng-galfit.fig5abc.eps, cpeng-galfit.fig5d.eps] {{\it Top}:
    Decomposition of NGC 221 (M32).  ({\it a}) The original image, ({\it b})
    and ({\it c}) residuals from GALFIT in positive greyscale.  The contour
    interval is arbitrary. {\it Bottom}: Isophotal analysis of the data image.
    The long-dashed lines are profiles of individual components, assuming a
    common center.  The solid line is the net sum of the sub-components.  Both
    the solid and long-dashed lines are seeing-removed profiles.  The
    short-dashed line that runs through the profile data points is a Nuker fit
    to the {\it observed} profile.  The region interior to the vertical dotted
    line is affected by seeing.
}
\end{figure}


\vfill\eject

\begin{figure} 
    \hbox{
        \hskip -0.4truein
        \includegraphics[scale=2.0,angle=90]{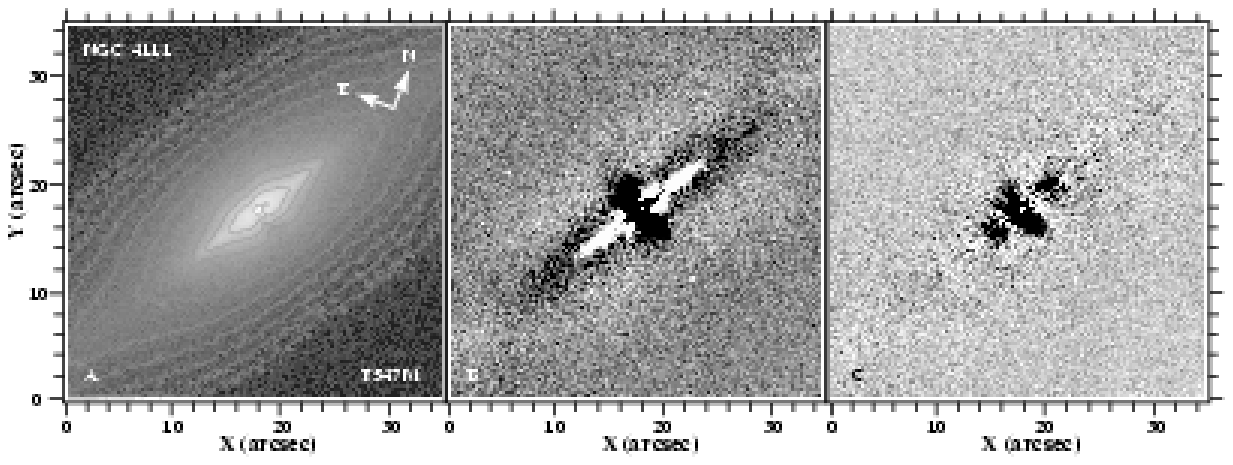} 
        \raisebox{1.5truein}{
        \includegraphics[width=4.0truein,height=7.truein]{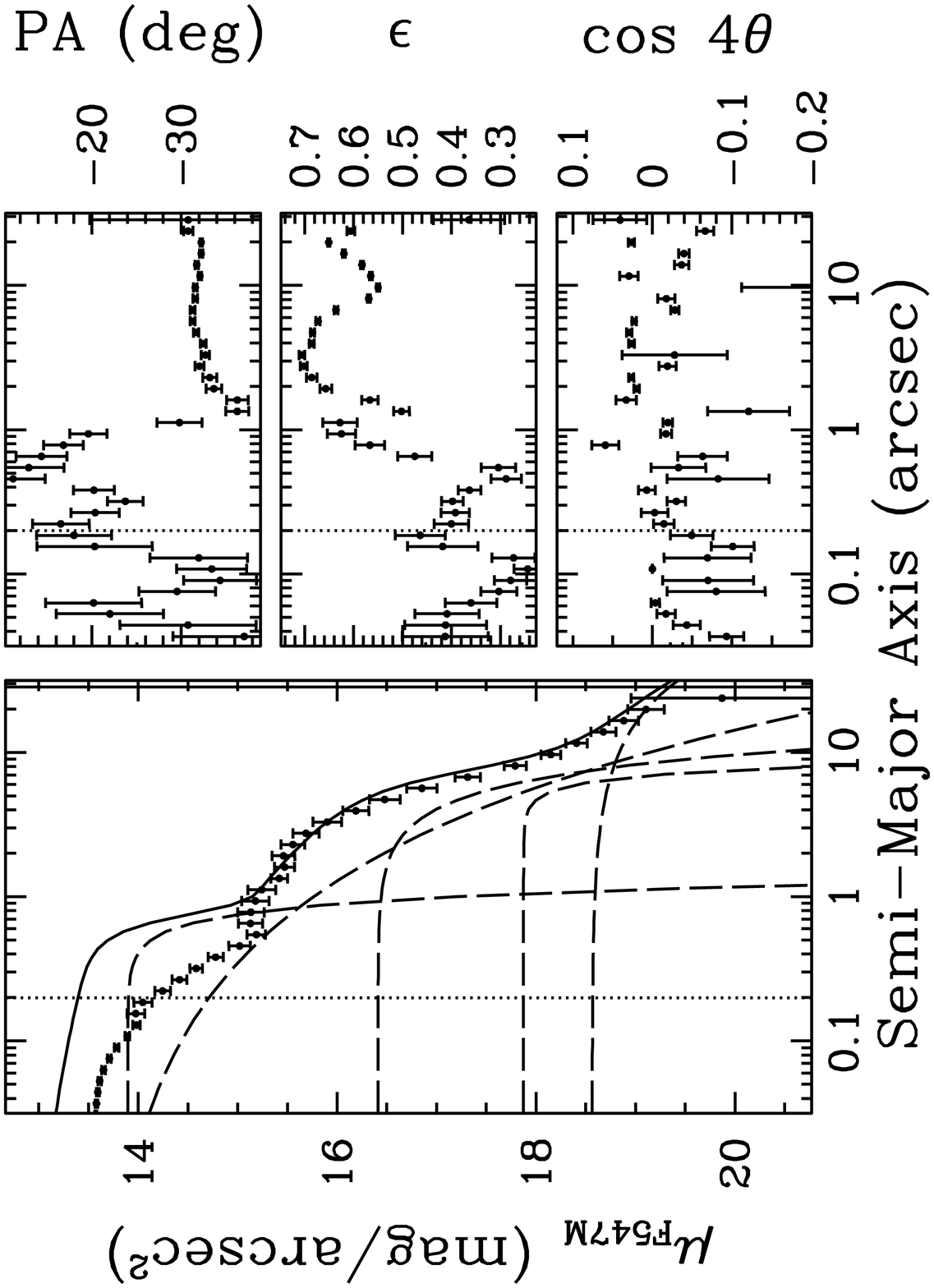}}
    }

    \figcaption[cpeng-galfit.fig7abc.eps, cpeng-galfit.fig7d.eps]
    {Decomposition of NGC 4111; see caption for Figure 5.}
\end{figure}

\vfill\eject

\begin{figure}
    \hbox{
        \hskip -0.4truein
        \includegraphics[scale=2.0,angle=90]{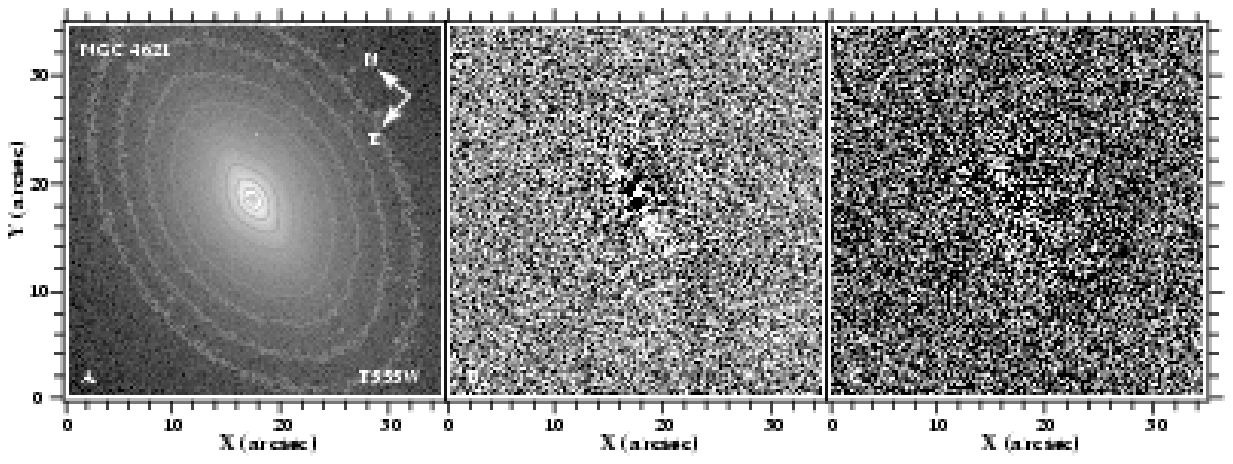} 
        \raisebox{1.5truein}{
        \includegraphics[width=4.0truein,height=7.truein]{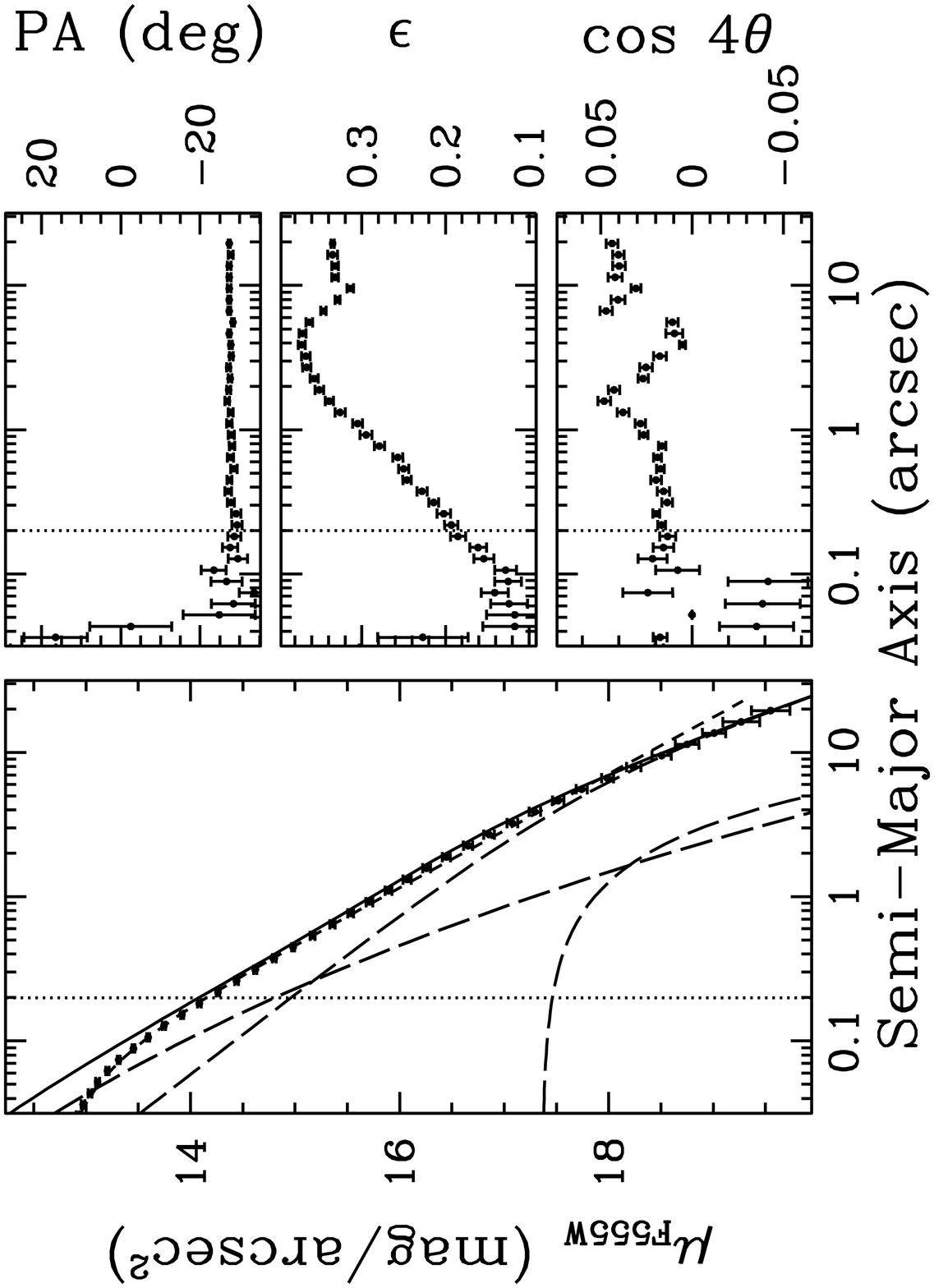}}
    }

    \figcaption[cpeng-galfit.fig8abc.eps, cpeng-galfit.fig8d.eps]
    {Decomposition of NGC 4621; see caption for Figure 5.}
\end{figure}


\vfill\eject

\begin{figure} 
    \hbox{
        \hskip -0.4truein
        \includegraphics[scale=2.0,angle=90]{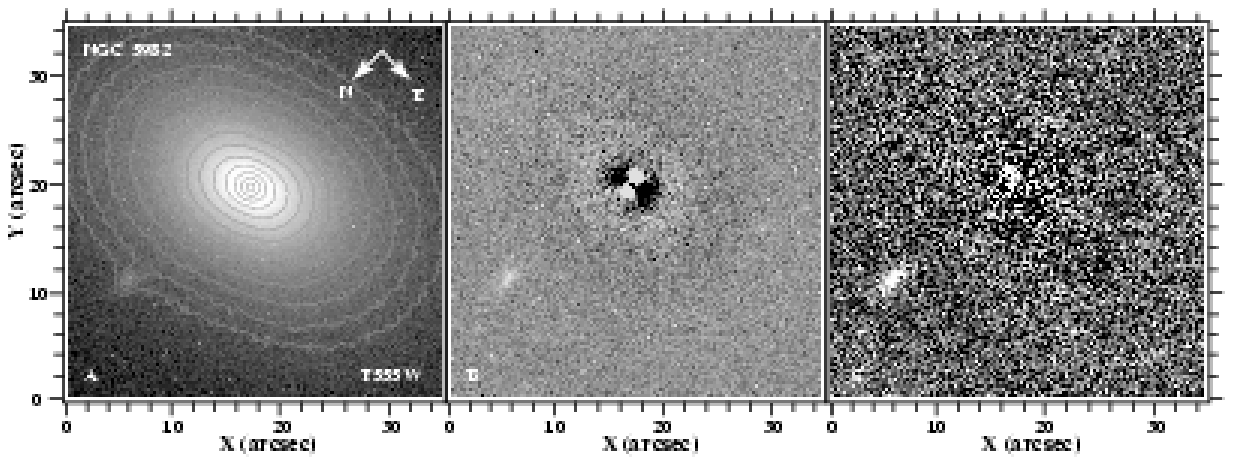} 
        \raisebox{1.5truein}{
        \includegraphics[width=4.0truein,height=7.truein]{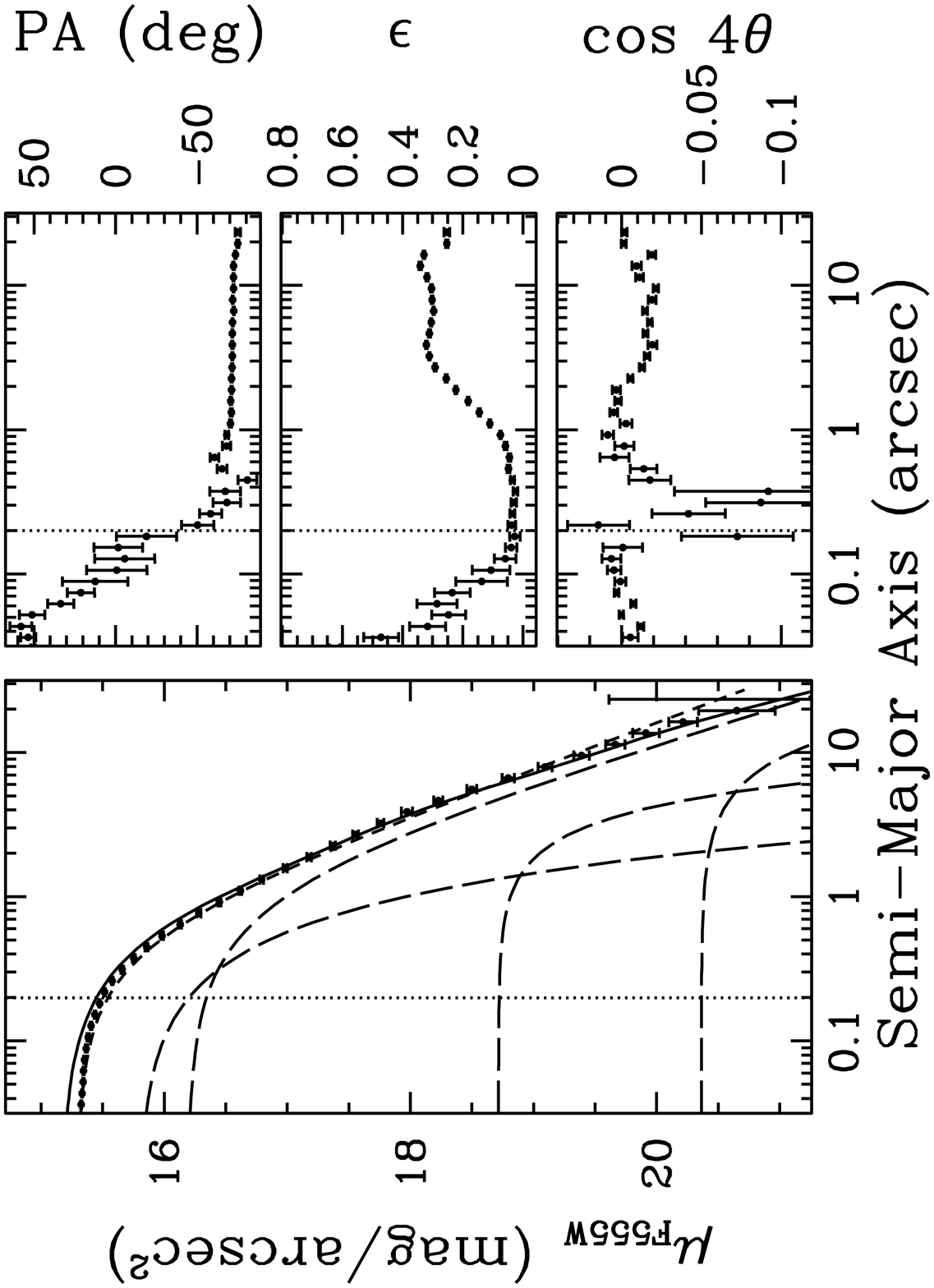}}
    }

    \figcaption[cpeng-galfit.fig10abc.eps, cpeng-galfit.fig10d.eps]
    {Decomposition of NGC 5982; see caption for Figure 5.}
\end{figure}


\vfill\eject

\begin{figure}
    \hbox{
        \hskip -0.4truein
        \includegraphics[scale=2.0,angle=90]{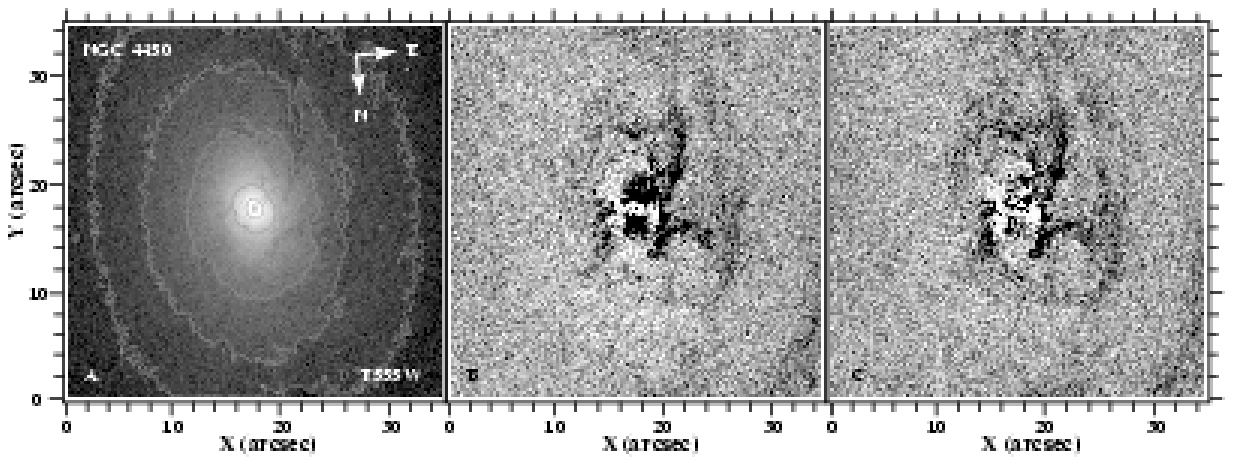} 
        \raisebox{1.5truein}{
        \includegraphics[width=4.0truein,height=7.truein]{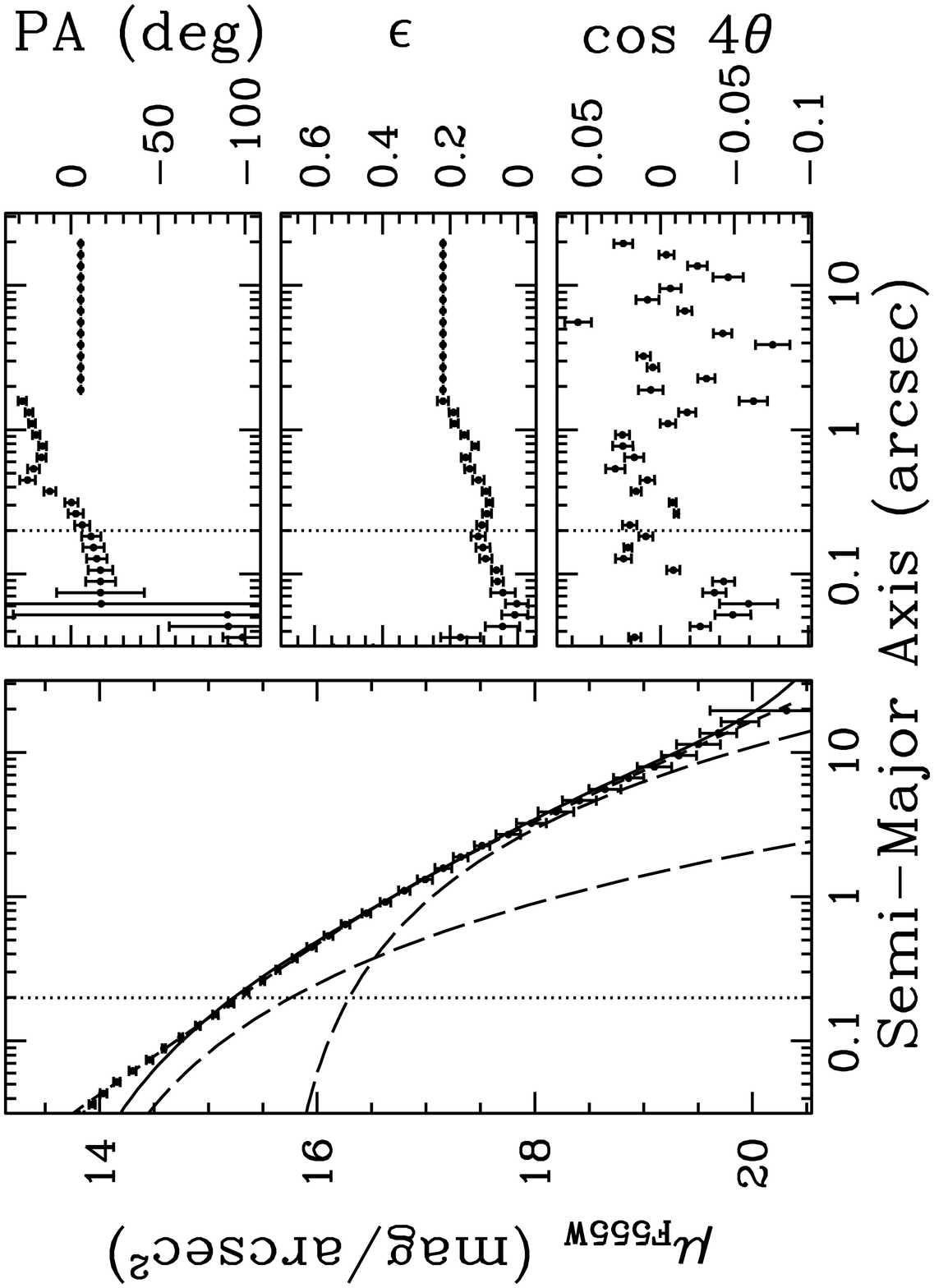}}
    }

    \figcaption[cpeng-galfit.fig12abc.eps, cpeng-galfit.fig12d.eps]
    {Decomposition of NGC 4450; see caption for Figure 5.}
\end{figure}

\vfill\eject

\begin{figure}
    \hbox{
        \hskip -0.4truein
        \includegraphics[scale=2.0,angle=90]{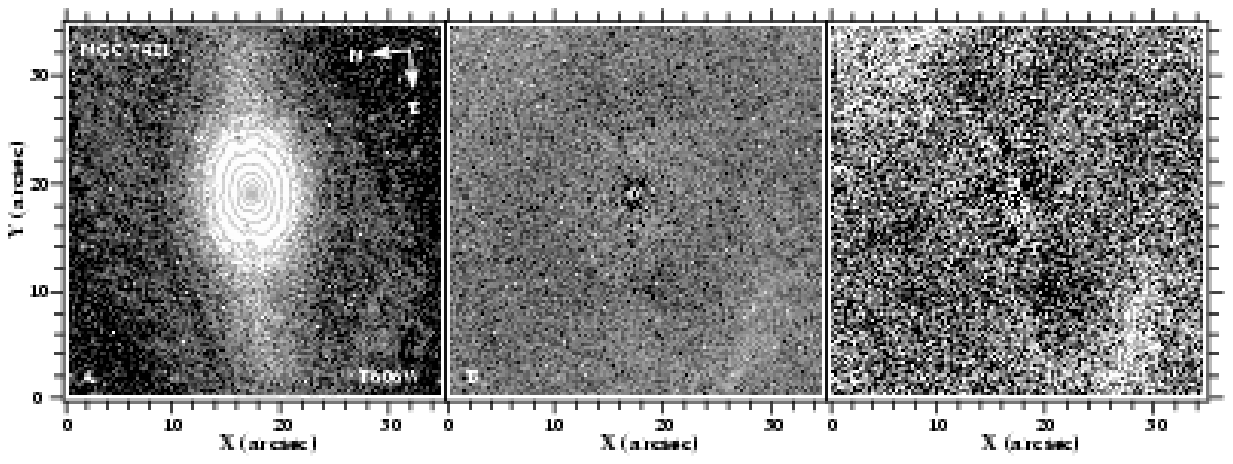} 
        \raisebox{1.5truein}{
        \includegraphics[width=4.0truein,height=7.truein]{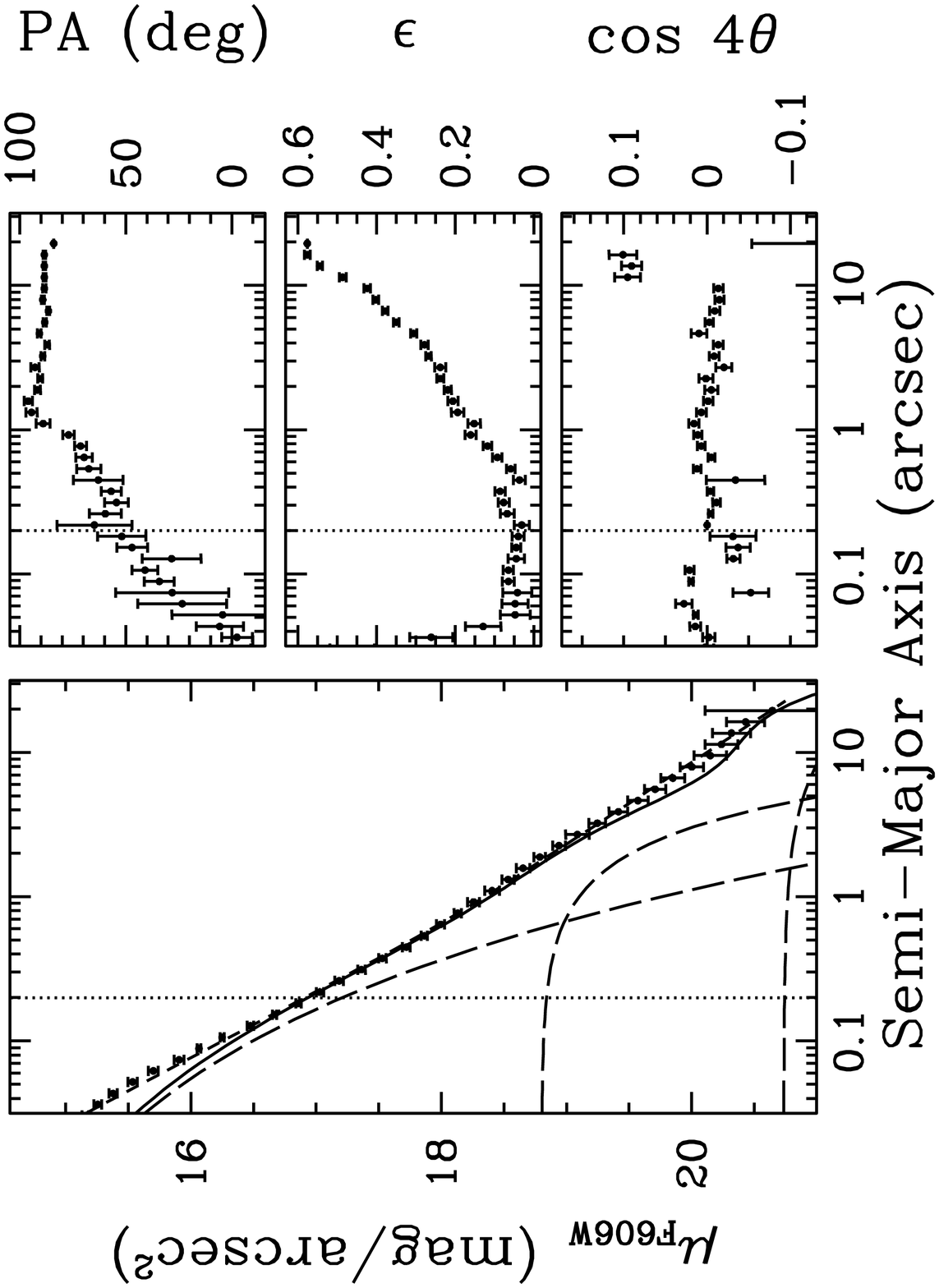}}
    }

    \figcaption[cpeng-galfit.fig13abc.eps, cpeng-galfit.fig13d.eps]
    {Decomposition of NGC 7421; see caption for Figure 5.}
\end{figure}

\vfill\eject

\begin{figure}[h]
    \plotone{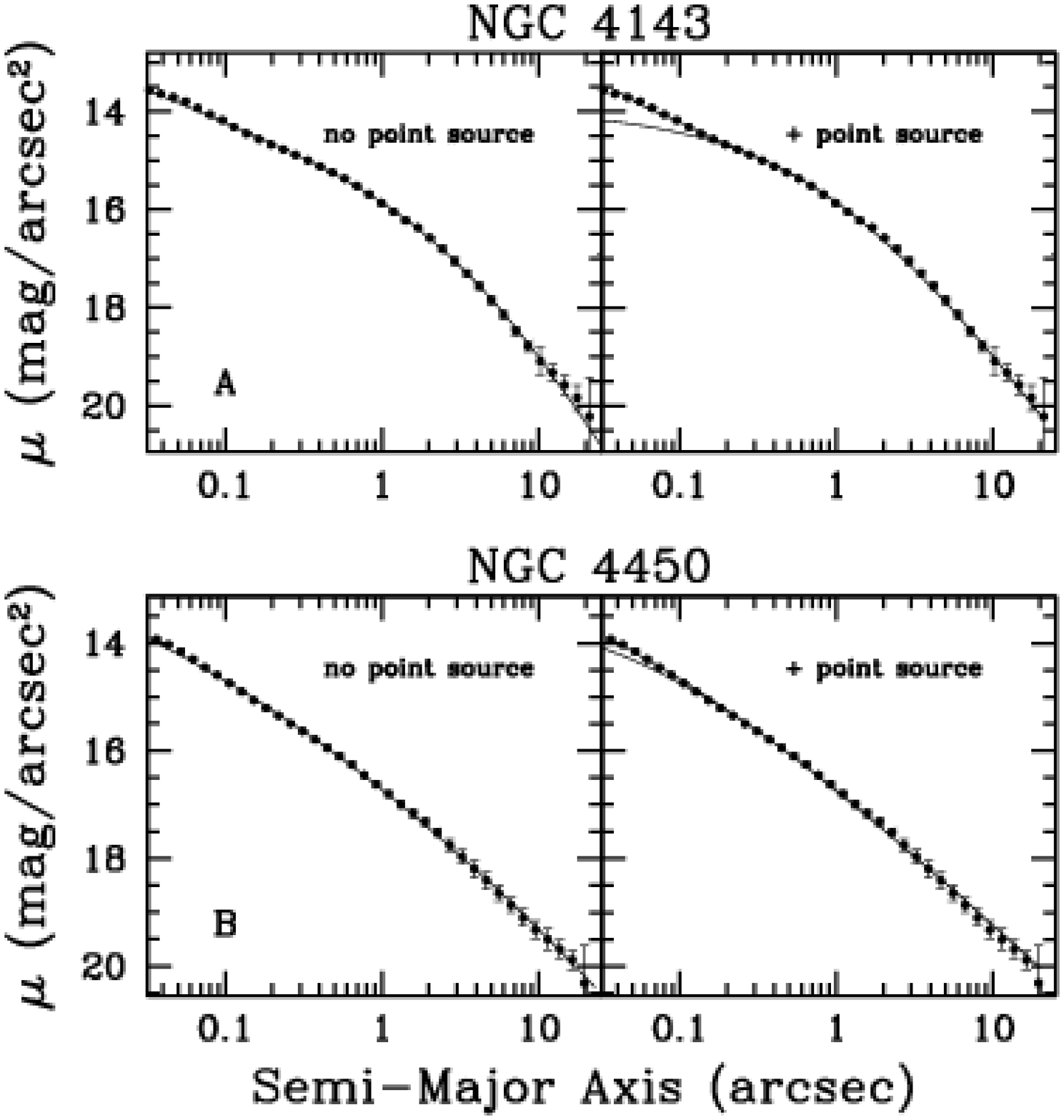}

    \figcaption[cpeng-galfit.fig14.eps]{Examples of ambiguities in 1-D
    decomposition of galaxies with point sources.  Round points are data and
    their error bars; solid lines are fits to the profiles using a Nuker
    function.}
\end{figure}

\vfill\eject

\begin{figure}[h]
    \plottwo {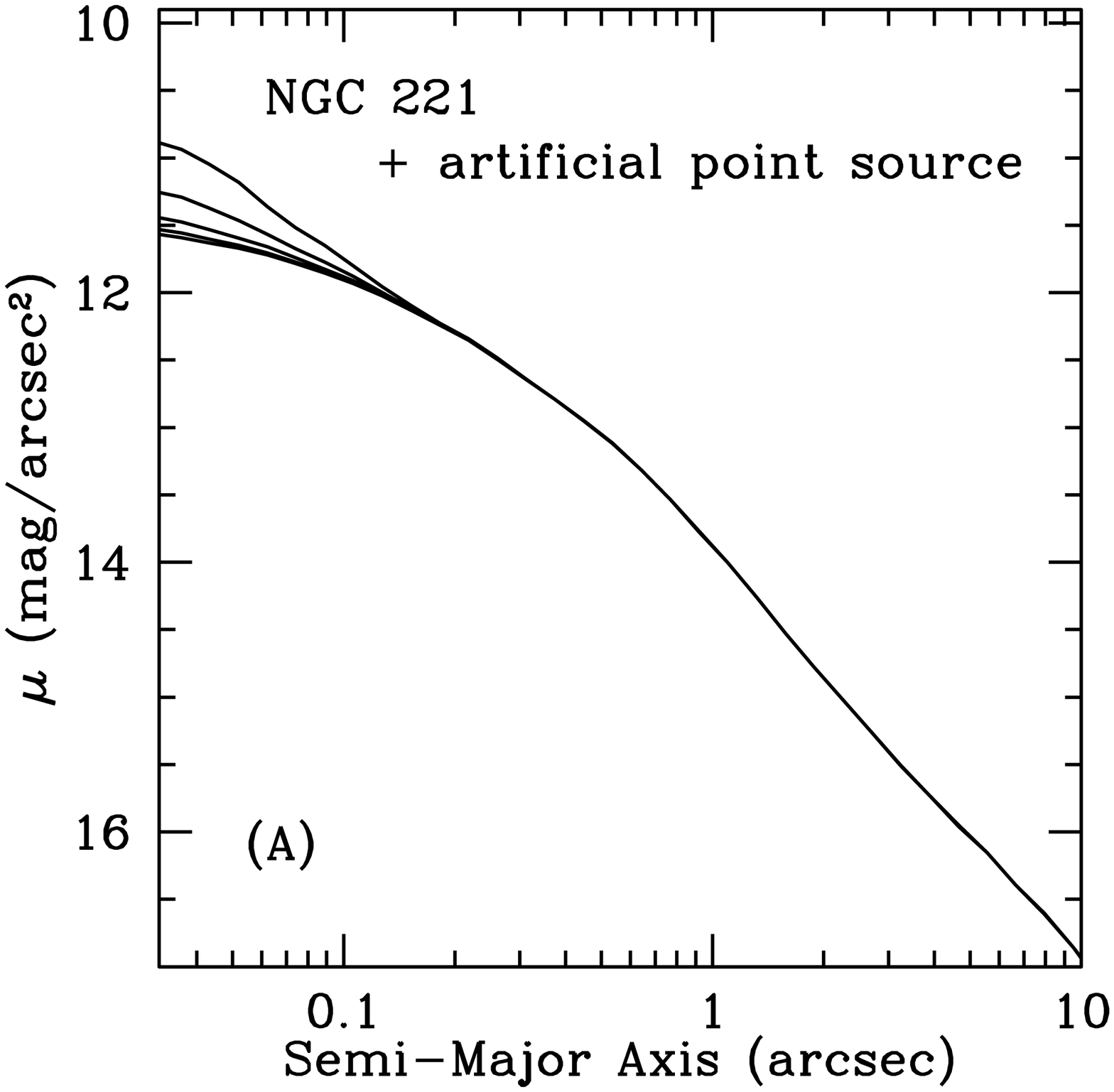}{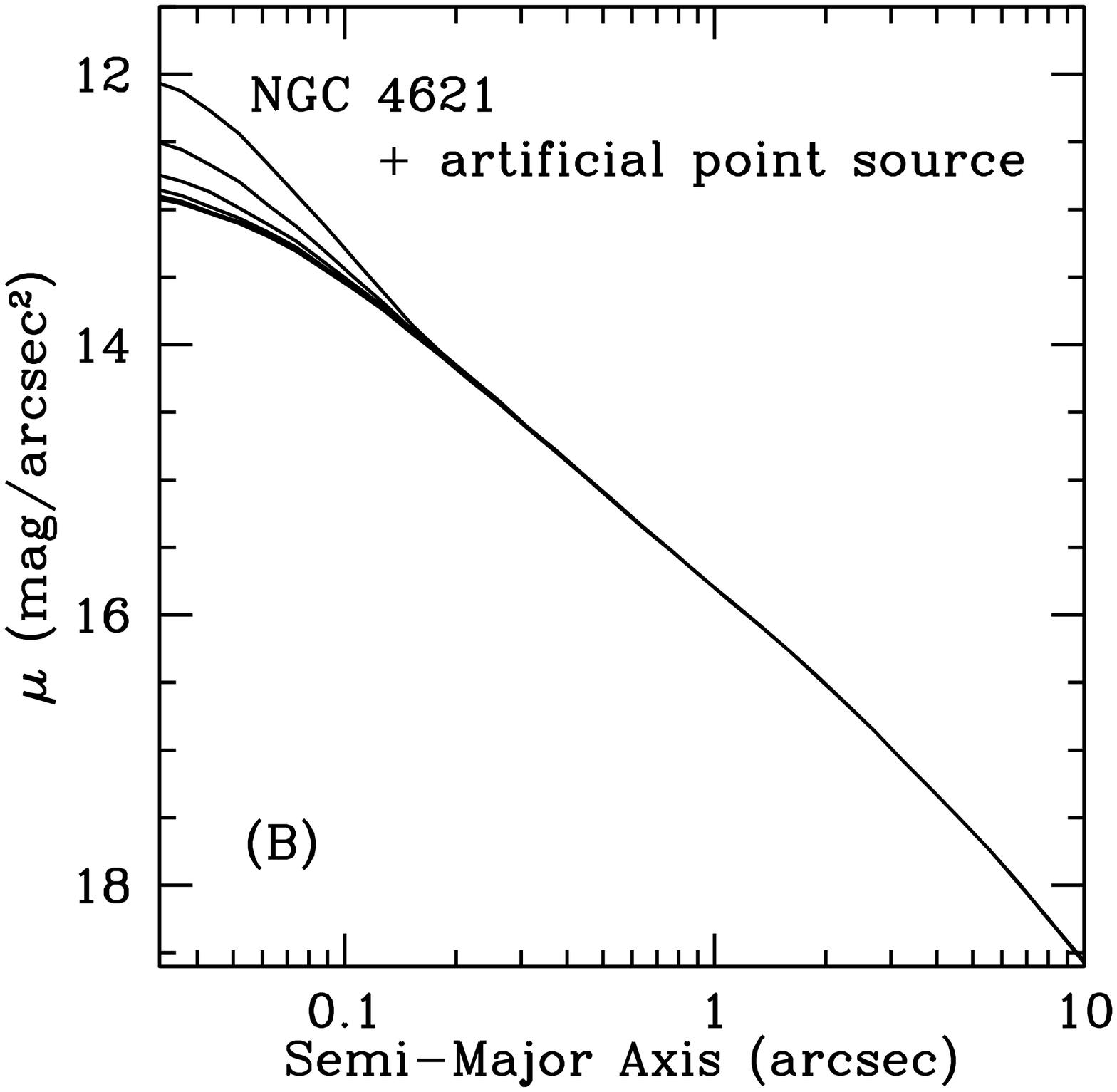}
    \bigskip
    \bigskip
    \bigskip
    \plotone {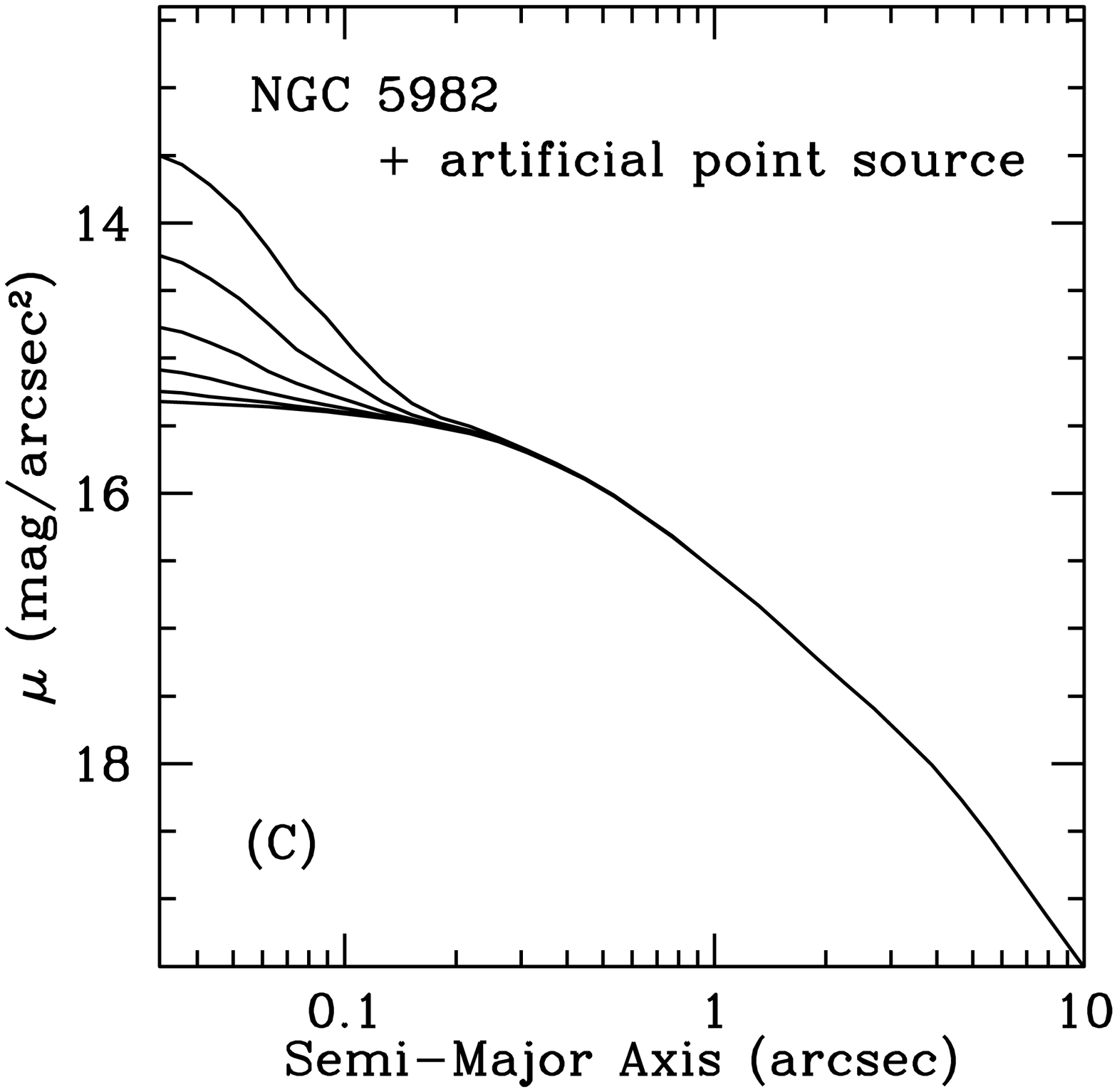}

    \figcaption[cpeng-galfit.fig15a.eps, cpeng-galfit.fig15b.eps,
    cpeng-galfit.fig15c.eps] {The surface brightness profile of ({\it a}) NGC
    221, ({\it b}) NGC 4621, and ({\it c}) NGC 5982 after adding a point
    source.  The sources added are 2, 3, 4, 5, 6, and 7 magnitudes fainter
    than the bulge, defined to be the region $r<0\farcs5$.}
\end{figure}

\vfill\eject

\begin{figure}
    \plotone {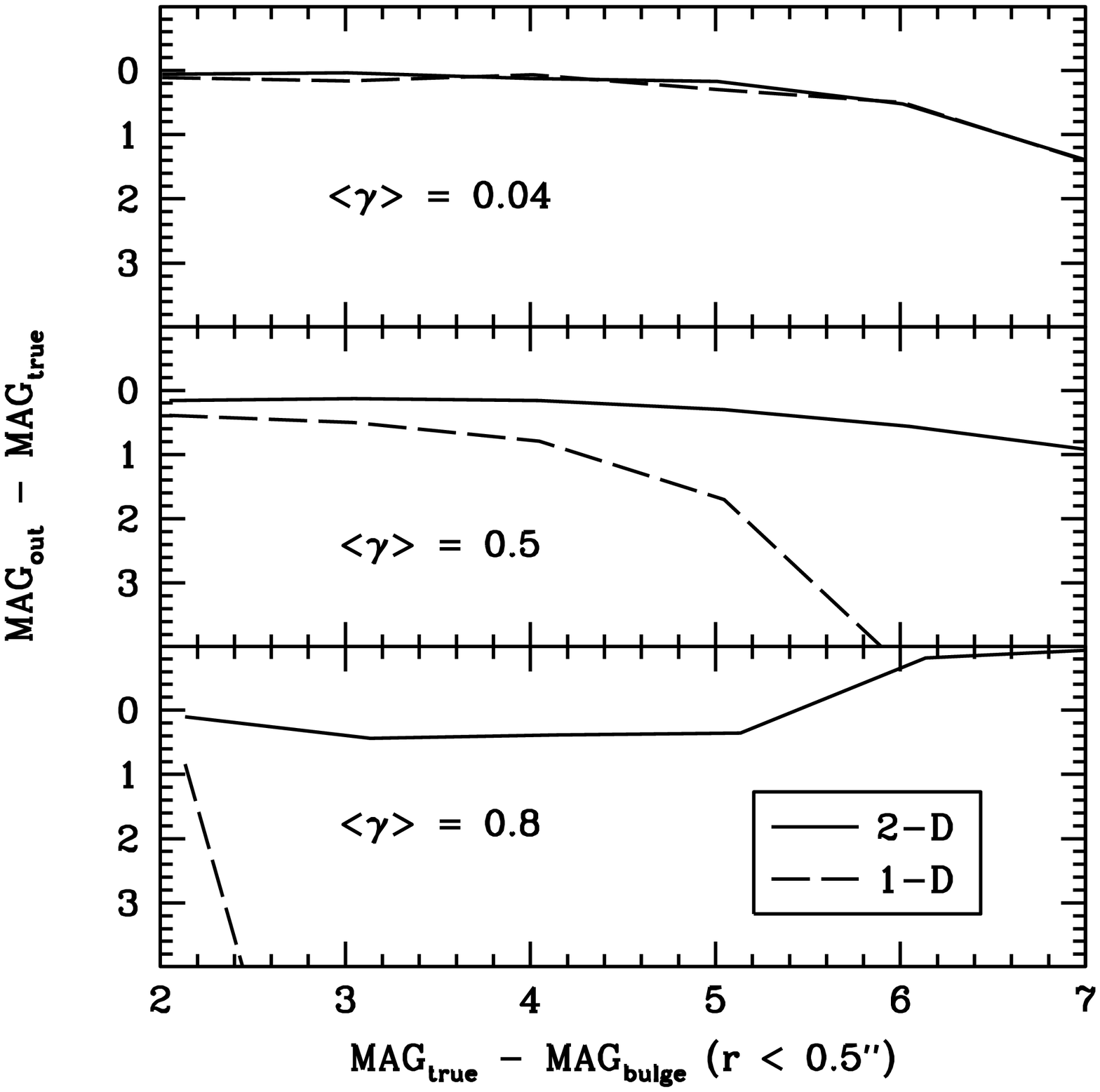}

    \figcaption[cpeng-galfit.fig16.eps]{Comparison between 1-D and 2-D
    decompositions for three different power-law cusps.  The ordinate gives
    the difference between input (MAG$_{\rm true}$) and recovered magnitudes
    (MAG$_{\rm out}$) of the point source, while the abscissa gives the input
    magnitude of the point source, normalized to the magnitude of the bulge
    (MAG$_{\rm bulge}$), defined to be the region $r < 0\farcs5$.  The nuclear
    cusp slope is defined by $\left<\gamma\right> \equiv \mbox{dlog} (I)/
    \mbox{dlog} (r)$ within $r < 0\farcs1$.}
\end{figure}

\vfill\eject

\begin{figure}
    \hbox{
        \hskip -0.4truein
        \includegraphics[scale=2.0,angle=90]{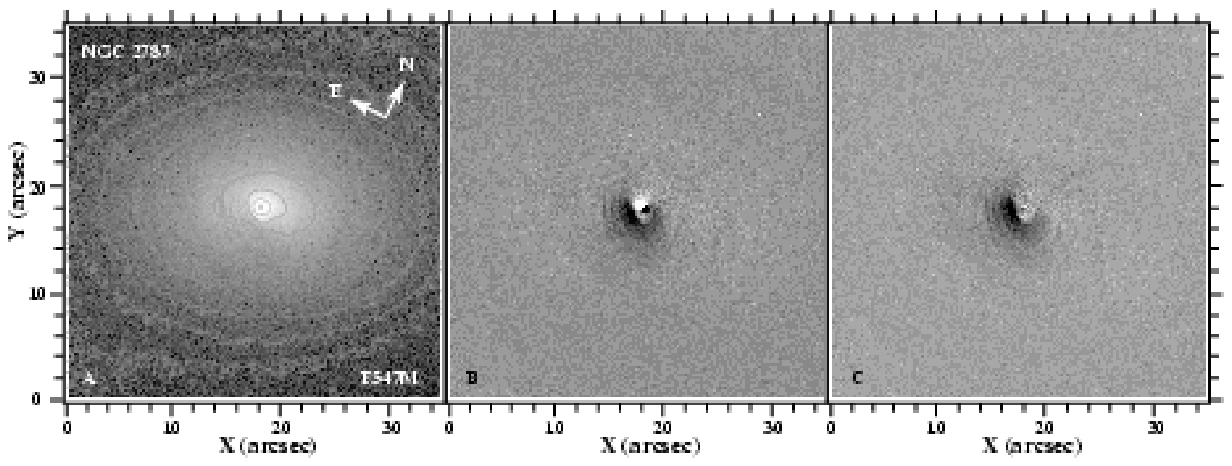} 
        \raisebox{1.5truein}{
        \includegraphics[width=4.0truein,height=7.truein]{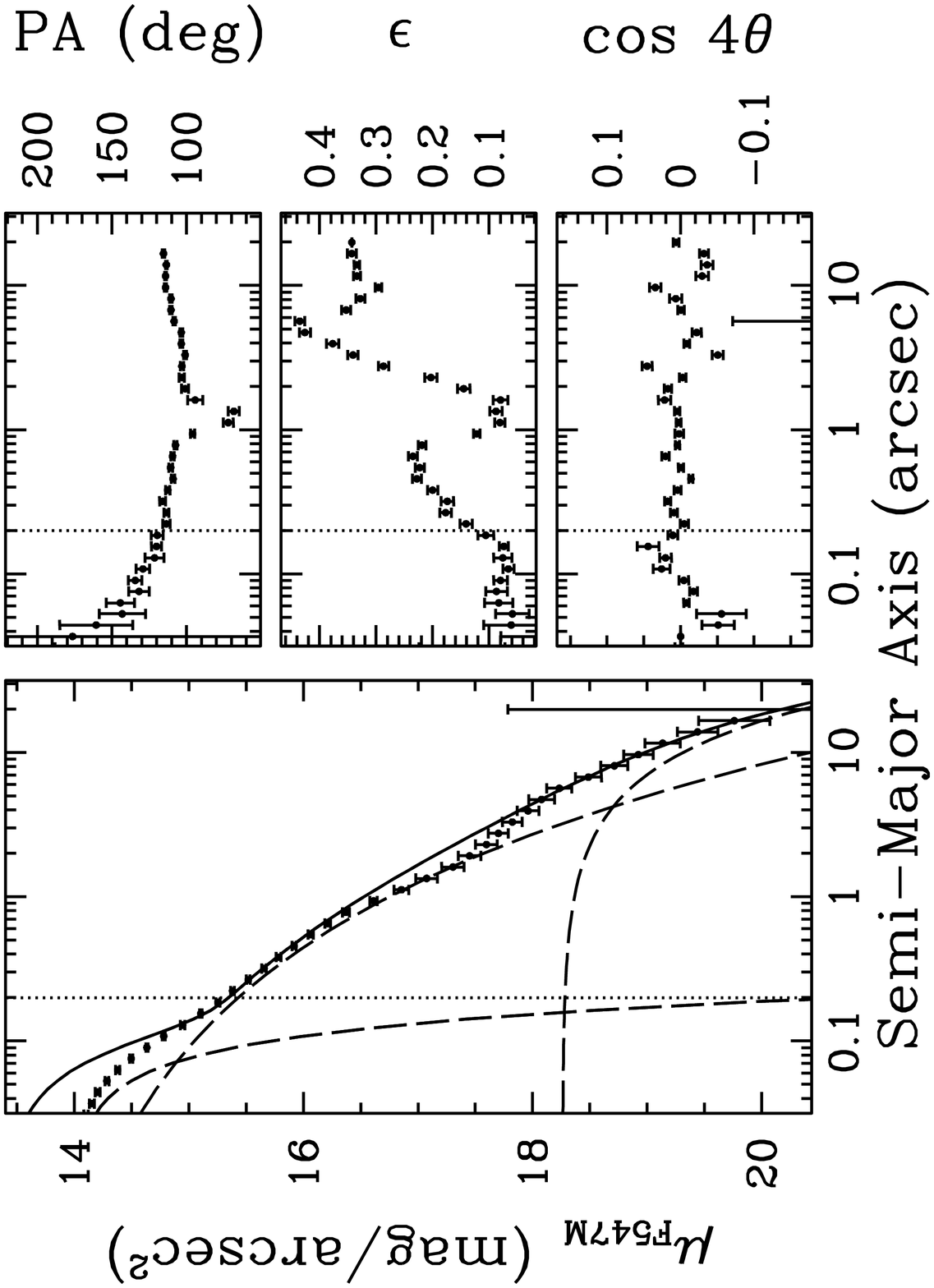}}
    }

    \figcaption[cpeng-galfit.fig17abc.eps, cpeng-galfit.fig17d.eps]
    {Decomposition of NGC 2787; see caption for Figure 5.}
\end{figure}

\end {document}